\documentclass[secnumarabic,12pt,tightenlines,eqsecnum,floats,aps,amsmath,amssymb,nofootinbib,superscriptaddress,showpacs,prd]{revtex4-1}

\usepackage{amsmath,amsthm,amssymb,amsfonts}
\usepackage{graphicx}
\usepackage{palatino}
\usepackage{supertabular}
\usepackage[mathcal]{euler}
\usepackage{mathcomp}
\usepackage{float}

\numberwithin{equation}{section}

\begin{document}

\title{Statistical description of the black hole degeneracy spectrum}

\author{J. Fernando \surname{Barbero G.}}
\affiliation{Instituto de
Estructura de la Materia, CSIC, Serrano 123, 28006 Madrid, Spain}

\author{Eduardo J. \surname{S. Villase\~nor}}
\affiliation{Instituto Gregorio
Mill\'an, Grupo de Modelizaci\'on y Simulaci\'on Num\'erica,
Universidad Carlos III de Madrid, Avda. de la Universidad 30, 28911
Legan\'es, Spain} \affiliation{Instituto de Estructura de la
Materia, CSIC, Serrano 123, 28006 Madrid, Spain}

\date{January 18, 2011}

\begin{abstract}
We use mathematical methods based on generating functions to study the statistical properties of the black hole degeneracy spectrum in loop quantum gravity. In particular we will study the persistence of the observed effective quantization of the entropy as a function of the horizon area. We will show that this quantization disappears as the area increases despite the existence of black hole configurations with a large degeneracy. The methods that we describe here can be adapted to the study of the statistical properties of the black hole degeneracy spectrum for all the existing proposals to define black hole entropy in loop quantum gravity.
\end{abstract}


\maketitle

\tableofcontents

\vfill\eject

\section{Introduction}

The study of black hole (BH) entropy within the framework provided by loop quantum gravity (LQG) is an interesting issue that illuminates important aspects of quantum gravity. The modeling of black holes by using space-times admitting isolated horizons as inner boundaries, and the subsequent quantization of this sector of general relativity, has been extensively explained in the literature \cite{Ashtekar:1997yu,Ashtekar:2000eq,Engle:2009vc,Engle:2010kt}. The resulting description provides a clear identification of the quantum BH degrees of freedom so that the standard quantum statistical definition of the entropy can be used.

For small black holes the detailed behavior of the entropy as a function of the horizon area has been explored in \cite{Corichi:2006wn,Corichi:2006bs,Agullo:2008yv}. A striking observation made in these papers is the fact that, in addition to the expected linear growth, the entropy displays a distinct staircase structure that amounts to its effective quantization. This is surprising because the spectrum of the area operator is not equally spaced. A detailed study of this phenomenon has been undertaken by resorting to combinatorial methods --in particular the use of generating functions-- and number theoretic ideas. These have been described in \cite{Agullo:2008eg,Agullo:2008yv,BarberoG.:2008ue,Agullo:2010zz}.

\begin{figure}[htbp]
\includegraphics[width=16.5cm]{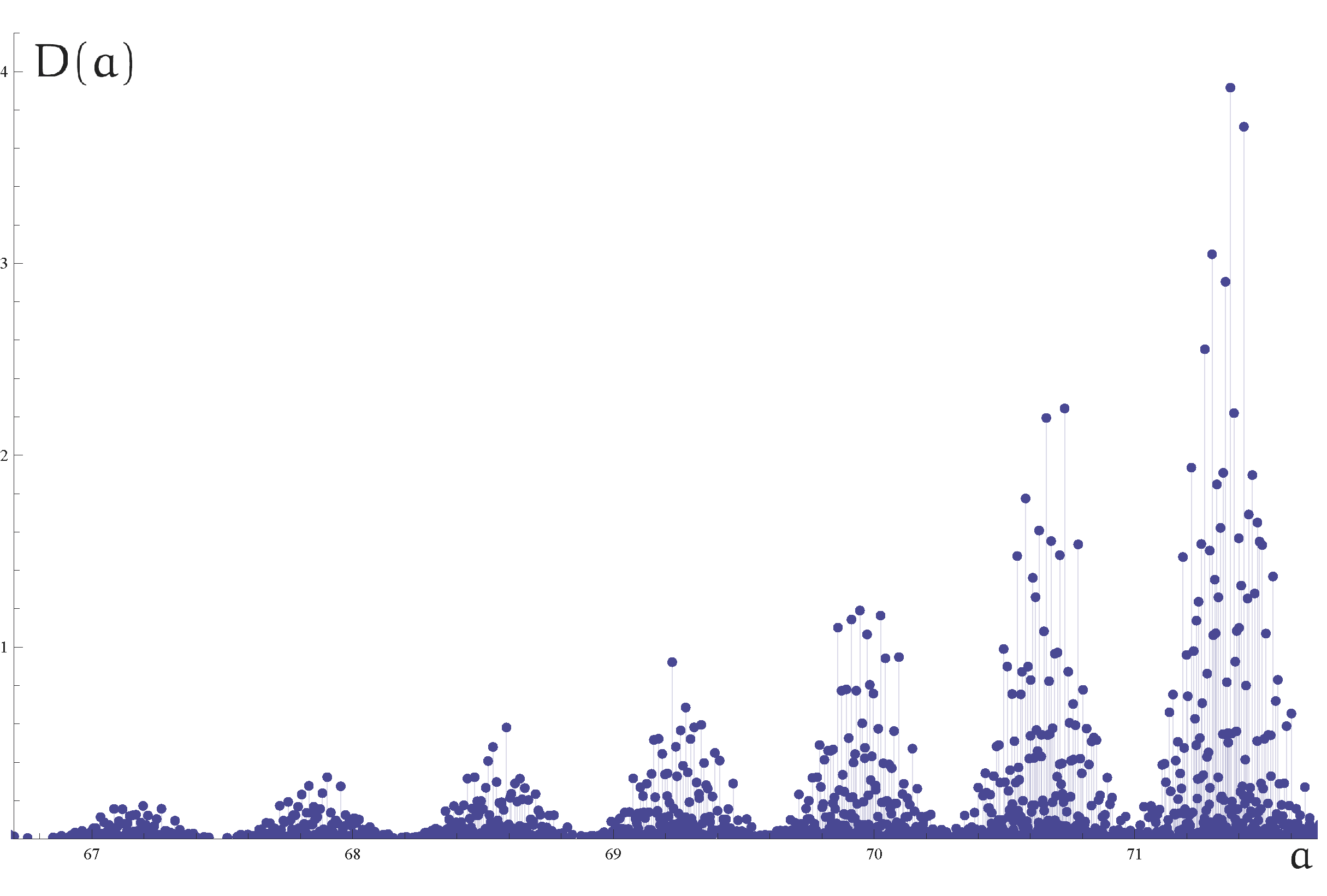}
\caption{Plot of the black hole degeneracy spectrum $D(a)$, in units of $10^{19}$, for a range of area values (in units of $4\pi\gamma\ell^2_P$, where $\gamma$ is the Immirzi parameter and $\ell_P$ the Planck length). The band structure can be traced all the way back to the smaller values of the area.} \label{Fig:intro}
\end{figure}

The so called black hole degeneracy spectrum is a way to encode the detailed information about BH configurations and their contributions to the entropy. In effect, the entropy can be computed as the integral of the black hole degeneracy distribution \cite{Corichi:2006wn,Agullo:2010zz}. When this picture is used, the results on the entropy quantization manifest themselves as a distinct peak structure in the degeneracy spectrum (see Fig. \ref{Fig:intro}). This fact led to the identification in reference \cite{Agullo:2008eg} of a peak counter --a function of the punctures of the spin network describing a BH state at the horizon-- that efficiently labels the configurations contributing to a given peak. An alternative way to do this has been given in \cite{Agullo:2010zz} as well as a generating function that singles out peak configurations. The main goal of the present paper is to use this master generating function to derive some important \textit{statistical} information about the peaks in the degeneracy spectrum and discuss its physical implications. The reason why we follow a statistical approach is the fact that an inspection of the nature of the degeneracy spectrum shows a combination of a simple coarse grained structure and a complicated detailed behavior as can be seen in Fig. \ref{Fig:intro}.

An important feature of our approach is the use of very strong results in Combinatorics that show a particular type of convergence to a Gaussian model when certain subsets of BH configurations are chosen. The obtention of the relevant statistical parameters, the mean and the variance, can be efficiently done in terms of the above mentioned generating functions. We want to point out that some of the methods that we will use in the paper are particular applications of general theorems in Combinatorics (see the excellent book by Flajolet and Sedgewick \cite{Flajolet}) suggesting that some features in the behavior of the entropy --in particular its effective quantization for small areas-- are actually of a very generic nature. This is also supported by the fact that this phenomenon has been seen in all the different proposals found in the literature and is insensitive to the implementation of the so called projection constraint \cite{Agullo:2010zz}. We want to mention here that we restrict our analysis in the main body of the paper to the prescription given by Domagala and Lewandowski in \cite{Domagala:2004jt} to compute the entropy. In any case, we will show that our methods can be easily adapted to deal with the other countings appearing in the literature, in particular the $SU(2)$ proposal of \cite{Engle:2009vc, Engle:2010kt} (see also \cite{Ghosh:2006ph,Kaul:1998xv}), and our conclusions extended to these cases.

The statistical information that we obtain has a direct physical application. By smoothing out the peaks in the degeneracy spectrum (or, rather, the steps in the entropy) and describing them as Gaussians (or better by Gaussian distributions, that can be written in terms of the error function $\mathrm{erf}$) it is possible to obtain a smoothed representation for the black hole entropy. The low area behavior, that has been studied so far in the literature, is captured in a very effective way by this model. It is possible to show that the interesting structure of the entropy seen for small black holes disappears in an area regime for which the smooth approximation is still valid. However, the preceding analysis does not exclude a revival of the entropy quantization for larger areas (or in the asymptotic limit) because the smoothed model fails to reproduce the exact value of the Immirzi parameter $\gamma$ and, hence, the correct growth rate of the entropy (although very good approximations for $\gamma$ are obtained in practice).

The lay out of the paper is the following. After this introduction we will devote Section \ref{basicdefinitions} to give the basic definitions related to the entropy and the black hole degeneracy spectrum. In Section \ref{stats} we will study the statistical properties of the peaks by introducing their moment-generating function. We will obtain the mean and the variance for the peak distribution and discuss the computation of higher moments. The approximation obtained by modeling the steps as Heaviside step functions (with discontinuities located at the area values given by the mean value of the areas associated with the peaks) will be discussed next in Section \ref{staircase}. As we will see, the approximation obtained in this way reproduces the behavior of the entropy for small areas remarkably well, though it is not suitable to understand the origin of the staircase structure itself. This can be better done by taking into account not only the mean but also the variance. We devote Section \ref{gaussian} to this issue. As we will see the low-area structure of the entropy can be neatly understood in this setting. Furthermore, we can also explain in quantitative terms how this structure disappears when the area increases.

The basic approach discussed in the first part of the paper can be improved in several ways. One of them consists in further partitioning the space of black hole configurations by introducing extra peak counters. A particularly simple description can be found by using two of them. This will allow us to explain, at least in the limit of small areas, the appearance of discrete substructures in the peaks of the BH degeneracy spectrum. The details of this are described in Section \ref{sect:substructure}. We also give there a quantitative comparison of the peak counter found in \cite{Agullo:2008eg} with other possible choices and conclude that it is the best one. We end the paper  in Section \ref{conclusions} with our conclusions and some details relevant for the extension of our methods to the $SU(2)$ formulation of \cite{Engle:2009vc}. A number of technical issues are left for the appendices. If not stated otherwise, areas in the paper will be given in units of $4\pi\gamma \ell_P^2$.

\section{Black hole entropy: Basic definitions}\label{basicdefinitions}

As we have mentioned in the introduction, some details in the entropy behavior as a function of the area are insensitive to the counting scheme that one chooses to follow (within the family of LQG inspired models). For the sake of concreteness most of the computations and results presented in the paper correspond to the Domagala-Lewandowski (DL) implementation \cite{Domagala:2004jt} of the original proposal of Ashtekar, Baez, Corichi and Krasnov \cite{Ashtekar:1997yu,Ashtekar:2000eq}. However we will briefly discuss at the end of the paper the relevance of our results for the recent $SU(2)$ proposal of \cite{Engle:2009vc,Engle:2010kt}.

An extended discussion of the number-theoretic and combinatorial methods that we will employ here can be found in \cite{Agullo:2010zz}, in particular the notation and definitions that we use in the paper. Nonetheless, and for the benefit of the reader, we give here the basic definitions that will be used.

In the DL approach, the entropy $S(a)$ (respectively $S_*(a)$, when the so called \textit{projection constraint} is ignored) of a quantum horizon of classical area $a$ is given by
$$S(a) = \log (1+\mathfrak{N}(a))\,,\quad \textrm{(repectively }S_*(a) = \log (1+\mathfrak{N}_*(a))\textrm{)},$$
where $\mathfrak{N}(a)$ (respectively $\mathfrak{N}_*(a)$) is  the number of all the finite, arbitrarily long, sequences $(m_1,\ldots,m_N)$ of non-zero half integers, such that:
$$\sum_{I=1}^N m_I=0, \quad 2\sum_{I=1}^N\sqrt{|m_I|(|m_I|+1)}\leq a\quad
\textrm{(respectively }2\sum_{I=1}^N\sqrt{|m_I|(|m_I|+1)}\leq a\textrm{)}.$$
The condition $\sum_I m_I=0$ is known as the projection constraint. The computation of both $S(a)$ and $S_*(a)$ can be efficiently performed in terms of the sets $\mathcal{C}(a')$, $a'\leq a$, of the  allowed configurations  for each  area $a'=\sum_{i}q_i\sqrt{p_i}\in \mathrm{sp}(\hat{a}^{{\scriptscriptstyle{LQG}}})$ belonging to the spectrum of the LQG area operator. Here, as pointed out in \cite{Agullo:2008yv}, $q_i\in \mathbb{N}\cup\{0\}$ and $p_i$ are the square-free integers ($p_1=2$, $p_2=3$, $p_3=5$, $p_4=6$, etc.).  A configuration $c\in\mathcal{C}(a)$, as defined in \cite{Agullo:2010zz}, is a (finite) multiset  $c=\{(k,N_k)\}$ in which each integer $k\in \mathbb{N}$ appears $N_k$ times (with $N_k\in \mathbb{N}\cup\{0\}$).  The set $\mathcal{C}$ of \textit{all} possible BH configurations is defined as the union of the configurations corresponding to the different area values
$$\mathcal{C}:=\bigcup_{a\in \mathrm{sp}(\hat{a}^{{\scriptscriptstyle{LQG}}})}\mathcal{C}(a)\,.$$

There are several functions $f:\mathcal{C}\rightarrow \mathbb{R}$ defined on the space of configurations with  a clear physical interpretation that we will use extensively in the following:
\begin{itemize}
\item For any configuration $c=\{(k,N_k)\}$,  $N(c)=\sum_kN_k$ represents the number of punctures defined by a spin network piercing the horizon,  $K(c)=\sum_k kN_k$ is (twice) the total spin, and $A(c)=\sum_k \sqrt{k(k+2)} N_k$ the area of the BH induced by $c$. We will use also some other functions defined on $\mathcal{C}$ such as the ``peak counter'' $P(c):=3K(c)+2N(c)$\,. These functions satisfy the bound $P/3<A<2P/5$ or, equivalently, $5A/2<P<3A$.
\item As explained in \cite{Agullo:2010zz}, the degeneracy  $d(c)$ of a configuration $c\in \mathcal{C}(a)$ allows us to compute the number $D(a)=\sum_{c\in \mathcal{C}(a)} d(c)$  of arbitrarily long sequences $(m_1,\ldots,m_N)$ of non-zero half integers, such that:
$$\sum_{I=1}^N m_I=0, \quad 2\sum_{I=1}^N\sqrt{|m_I|(|m_I|+1)}= a.$$
In terms of the so called  BH degeneracy spectrum $D(a)$, the BH entropy is given by
$$
\exp S(a) =1+\sum_{a'\leq a} D(a')\,.
$$

\item When the projection constraint is ignored, the degeneracies   $d_*(c)$ of a configuration $c\in \mathcal{C}(a)$  give us the number  $D_*(a)=\sum_{c\in \mathcal{C}(a)}d_*(c)$  of arbitrarily long sequences $(m_1,\ldots,m_N)$ of non-zero half integers, such that:
$$2\sum_{I=1}^N\sqrt{|m_I|(|m_I|+1)}=a.$$
The BH degeneracy spectrum  $D_*(a)$ in the case when the projection constraint is not considered can be used to compute the value of the entropy
$$
\exp S_*(a) =1+\sum_{a'\leq a} D_*(a')\,.
$$
\end{itemize}
For a given area $a=q_1\sqrt{p_1}+q_2\sqrt{p_2}+\cdots$,  the values of $D(a)$ and $D_*(a)$ can be encoded in the coefficients of the  generating functions $G(z;x_1,x_2,\dots)$ and $G_*(x_1,x_2,\dots)$:
\begin{eqnarray*}
D(a)&=&[z^0][x_1^{q_1}x_2^{q_2}\cdots]G(z;x_1,x_2,\dots)\\&=&[z^0][x_1^{q_1}x_2^{q_2}\cdots]\left(1-\sum_{i=1}^\infty \sum_{ n=1}^\infty (z^{k_ n ^i}+z^{-k_ n ^i}) x_i^{y_ n^i}\right)^{-1}\,,\\
D_*(a)&=&[x_1^{q_1}x_2^{q_2}\cdots]G_*(x_1,x_2,\dots)\\
&=&[x_1^{q_1}x_2^{q_2}\cdots]\left(1-2\sum_{i=1}^\infty \sum_{ n=1}^\infty x_i^{y_ n^i}\right)^{-1}\,.
\end{eqnarray*}
For each square-free $p_i$, the terms $(z^{k_ n ^i}+z^{-k_ n ^i}) x_i^{y_ n^i}$ and $x_i^{y_ n^i}$ appearing in the corresponding generating functions are built from the solutions $\{(k^i_ n,y_ n^i)\}_ n$ to the Pell equations $(k+1)^2-p_iy^2=1$ (see \cite{Agullo:2010zz} for details). Here $[z^0][x_1^{q_1}x_2^{q_2}\cdots]G(z;x_1,x_2,\dots)$ denotes the coefficient of the $z^0 x_1^{q_1}x_2^{q_2}\cdots$ term in a Laurent expansion of $G(z;x_1,x_2,\dots)$ about $z=0$, $x_1=0,\ldots$

It is important to notice that the family  $\{\mathcal{C}(a)\subset \mathcal{C}\,:\, a\in \mathrm{sp}(\hat{a}^{{\scriptscriptstyle{LQG}}})\}$ provides us with a partition of the configuration space $\mathcal{C}$ defined in terms of the level sets of the area function $\mathcal{C}(a)=A^{-1}(a)$. If we are given any other function $P$ in the configuration space (in particular  $P(c)=3K(c)+2N(c)$) it is possible to define a different partition $\mathcal{C}=\bigcup_p \mathcal{P}_p$ using the level sets $\mathcal{P}_p=P^{-1}(p)$. This means that the sets $\mathcal{C}_p(a)=P^{-1}(p)\cap A^{-1}(a)$ define a finer partition than either $\bigcup_a\mathcal{C}(a)$ or $\bigcup_p \mathcal{P}_p$:
$$
\mathcal{C}=\bigcup_p \bigcup_a \mathcal{C}_p(a)\,.
$$
Notice that $\mathcal{C}(a)=\bigcup_p \mathcal{C}_p(a)$ and $\mathcal{P}_p=\bigcup_a\mathcal{C}_p(a)$. This fact can be used to compute the entropy as
\begin{eqnarray*}
\exp S(a) =1+\sum_p\sum_{a'\leq a} D(a'\,|\,p)\,,\quad \exp S_*(a) =1+\sum_p\sum_{a'\leq a} D_*(a'\,|\,p)
\end{eqnarray*}
where
$$
D(a\,|\,p)=\sum_{c\in \mathcal{C}_p(a)} d(c)\,,\quad D_*(a\,|\,p)=\sum_{c\in \mathcal{C}_p(a)} d_*(c)\,.
$$
This is so because
\begin{eqnarray}
\exp S(a) &=&1+\sum_{a'\leq a} D(a')=1+\sum_{a'\leq a} \sum_{c\in \mathcal{C}(a')} d(c)=1+\sum_{a'\leq a}\sum_p \sum_{c\in \mathcal{C}_p(a')} d(c)\nonumber\\
&=&1+\sum_p \sum_{a'\leq a} \sum_{c\in \mathcal{C}_p(a')} d(c)=1+\sum_p\sum_{a'\leq a} D(a'\,|\,p) \label{D}
\end{eqnarray}
and equivalently for $\exp S_*(a)$. Finally, it is important to notice that, when the partition is defined by the functions of the type $P(\alpha,\beta):=\alpha K+\beta N$ (a generalized ``linear'' counter with positive integer coefficients), the numbers $D(a\,|\,p)$ and $D_*(a\,|\,p)$ associated with an area $a=q_1\sqrt{p_1}+q_2\sqrt{p_2}+\cdots$ can be derived as
$$
D(a\,|\,p)=[z^0][x_1^{q_1}x_2^{q_2}\cdots ][\nu^p] G(\nu,z;x_1,x_2,\dots)\,,\quad D_*(a\,|\,p)=[x_1^{q_1}x_2^{q_2}\cdots ][\nu^p] G(\nu;x_1,x_2,\dots)\,,
$$
from the master BH generating functions
\begin{eqnarray}
 G(\nu,z;x_1,x_2,\dots)&:=&\left(1-\sum_{i=1}^\infty \sum_{ n=1}^\infty (z^{k_ n ^i}+z^{-k_ n ^i}) \nu^{\alpha k_ n ^i +\beta} x_i^{y_ n^i}\right)^{-1}\,,\label{master}\\
  G_*(\nu;x_1,x_2,\dots)&:=&\left(1-2\sum_{i=1}^\infty \sum_{ n=1}^\infty  \nu^{\alpha k_ n ^i +\beta} x_i^{y_ n^i}\right)^{-1}\,.\label{master*}
\end{eqnarray}
Notice that these generating functions are normalized in such a way that $D(0)=D(0\,|\,0)=D_*(0)=D_*(0\,|\,0)=1$, and $D(0\,|\,p)=D_*(0\,|\,p)=0$ for $p\neq 0$.

\section{Statistical properties of the peaks}\label{stats}

\begin{center}
\begin{figure}[htbp]
\includegraphics[width=16.5cm]{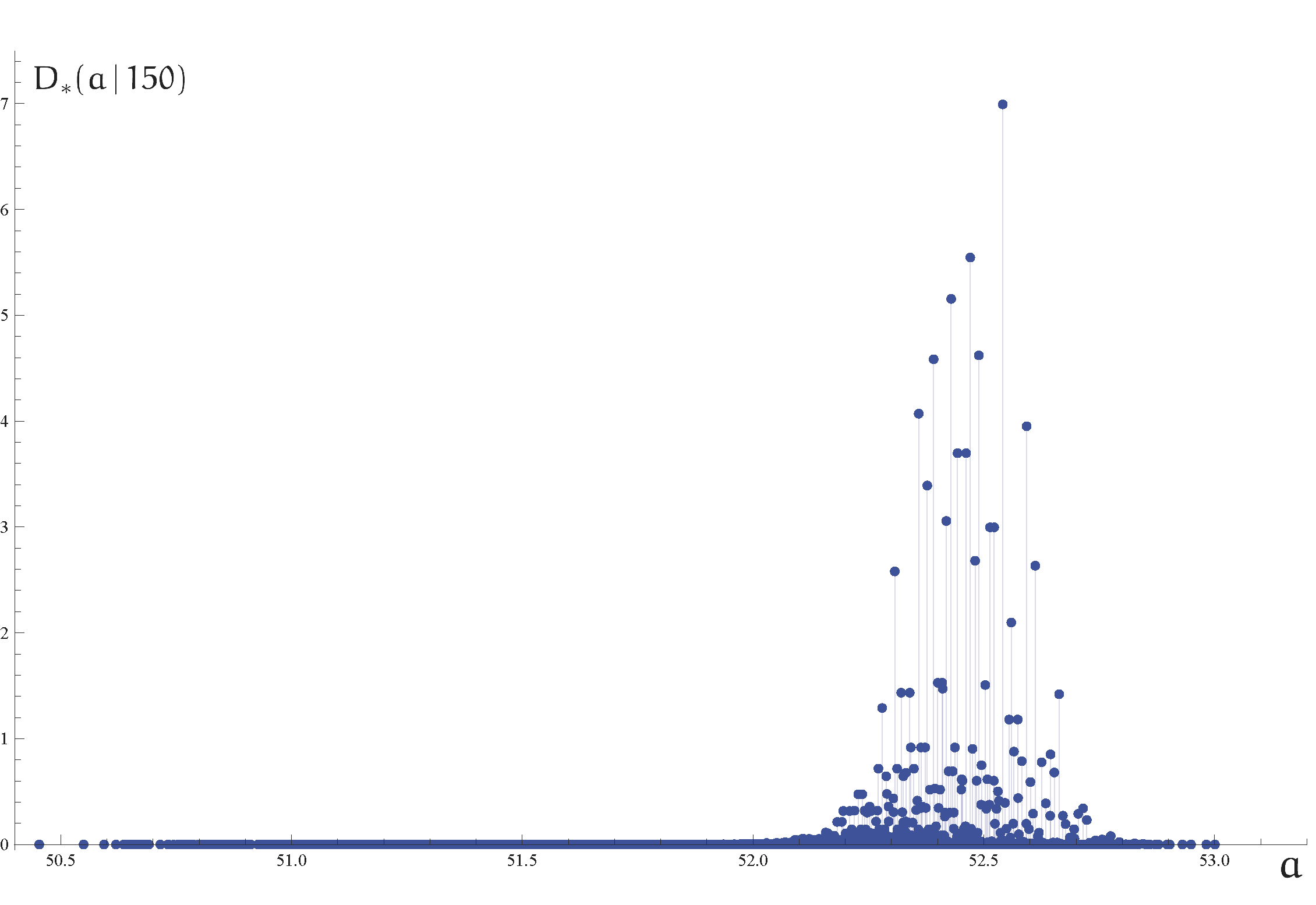}
\caption{Peak corresponding to $p=150$ in units of $10^{14}$.  By plotting the values of $D_*(a\,|\,p)$ one gets, for the largest degeneracy values, a roughly symmetric shape. The values corresponding to the largest degeneracies can be readily seen in the plot. Notice, however, that the peak extends to the left much farther than it does to the right. This phenomenon can be easily seen in a plot of the logarithm of $D_*(a\,|\,p)$ that also displays a distinct subpeak structure to be discussed later (see Fig. \ref{Fig:subpeaks}).} \label{Fig:peak}
\end{figure}
\end{center}

The starting point of our analysis is to introduce a convenient partition of the space of black hole configurations that is adapted to the description of the peak structure seen in Fig. \ref{Fig:intro} for the BH degeneracy spectrum (or, alternatively, to the steps of the entropy). This partition is performed by introducing the peak counter $P=3K+2N$ defined above. A \textit{peak} in the space of BH configurations $\mathcal{P}_p\subset \mathcal{C}$ is defined as consisting of those configurations $c$ corresponding to a pre-selected value $p$ of $P(c)$. We have then
$$
\mathcal{C}=\bigcup_{p} \mathcal{P}_p=\bigcup_{p} \bigcup_a \mathcal{C}_p(a)\,.
$$
For a fixed value of $p$ there are, of course, configurations corresponding to different values of the area (within a bounded range $P/3<A<2P/5$) and different degeneracies $d(c)$ (or $d_*(c)$); this is so because $\mathcal{P}_p=\bigcup_a \mathcal{C}_p(a)$. In fact, if one plots $D(a,p)=\sum_{c\in\mathcal{C}_p(a)}d(c)$  or $D_*(a,p)=\sum_{c\in\mathcal{C}_p(a)}d_*(c)$ as a function of $a$ (for a fixed value of $p$) one gets a regular structure with the form of peak in the black hole degeneracy, as shown in Fig. \ref{Fig:peak}.
It is obviously possible to reconstruct the degeneracy spectra that have already appeared in the literature \cite{Agullo:2010zz} by adding up the contributions of these peaks for all the values of $p$.
The regular shape seen in Figure \ref{Fig:peak} strongly suggests that a Gaussian approximation can provide a good description of the peaks. This fact leads naturally to the consideration of \textit{statistical methods} to study BH entropy. In fact, this is the main theme of this paper. We want to emphasize from the start that we do not merely compute statistical parameters (the mean, the variance and, eventually, higher moments) by fitting Gaussian profiles to the peak data but, rather, obtain them exactly from the BH generating functions. In other words, we will not just use descriptive statistics, but employ the very powerful analytical tools available for a wide class of combinatorial problems (involving generating functions of the same type as the ones that we use in this paper \cite{Flajolet}). This will allow us to make predictions regarding the statistical parameters of arbitrary peaks and use them to study the behavior of the BH entropy.

A statistical treatment requires us to give a weight to each configuration. In our problem this is naturally provided by the degeneracy $d(c)$ --or, respectively, $d_*(c)$. The relevant objects to be computed are the expectation values of the powers of the area (taken as a random variable) conditioned by a fixed value of $p$:
\begin{eqnarray}
E[A^n\,|\,p]&:=&\frac{\sum_{c\in\mathcal{P}_p}d(c)A^n(c)}{\sum_{c\in\mathcal{P}_p}d(c)}\,,\quad
E_*[A^n\,|\,p]:=\frac{\sum_{c\in\mathcal{P}_p}d_*(c)A^n(c)}{\sum_{c\in\mathcal{P}_p}d_*(c)}\,.
\label{expectation}
\end{eqnarray}
In the first case (where the projection constraint is taken into account) only even values of $p$ have be considered because $\sum_{c\in\mathcal{P}_p}d(c)$ is otherwise zero. As we will explain later, we will use the relevant moments defined by this formula to build a smooth approximation for the shape of each step in the entropy. This will require us to "de-normalize" the distribution by multiplying it by the total peak degeneracy $\sum_{c\in\mathcal{P}_p}d(c)$ (or $\sum_{c\in\mathcal{P}_p}d_*(c)$). The standard way to compute $E[A^n\,|\,p]$ and $E_*[A^n\,|\,p]$ relies on the use of the so called \textit{moment-generating function} associated with the random variable $A$. A remarkable feature of the combinatorial approach that we follow to study black hole entropy in LQG is the fact that this moment-generating function can be easily derived from the master generating functions (\ref{master}) or (\ref{master*}) given above. We will start by looking at the case where the projection constraint is ignored. The incorporation of the projection constraint will be discussed afterwards. Though this problem is more complicated, there are no important conceptual differences as far as our treatment is concerned.

\subsection{Moment-generating function: Ignoring the projection constraint}

Let us take as the starting point the master generating function $G_*(\nu;x_1,\ldots,)$, defined in (\ref{master*}), where the variable $x_i$ refers to the square free integer $p_i$. By substituting  $x_i=e^{-s\sqrt{p_i}}$, as is standard in this setting \cite{Agullo:2010zz}, we obtain
\begin{equation}
G_*(\nu,s):=G_*(\nu;e^{-s\sqrt{p_1}},e^{-s\sqrt{p_2}},\ldots)=\frac{1}{1-2\sum_{k=1}^\infty \nu^{3k+2}e^{-s\sqrt{k(k+2)}}}\,.
\label{genfunction}
\end{equation}
By construction, it is obvious \cite{Agullo:2010zz} that
$$
[\nu^p]G_*(\nu,s)=\sum_{c\in\mathcal{P}_p}d_*(c)e^{-s A(c)}\,
$$
and, hence,
\begin{equation}
E_*[\exp(-s A)\,|\,p]=\frac{[\nu^p]G_*(\nu,s)}{[\nu^p]G_*(\nu,0)}\,.
\label{momentgf}
\end{equation}
Modulo normalizing factors, and the exchange $s\mapsto -s$, the function $g(s\,|\,p):=[\nu^p]G_*(\nu,s)$ is the standard moment-generating function used in Mathematical Statistics and, hence, $\log g(s\,|\,p)$ is the cumulant-generating function used in Statistical Physics. Notice that our sign convention originates in the use of Laplace transforms to write down closed expressions for the black hole entropy \cite{Meissner:2004ju,G.:2008mj}. By computing the derivatives of (\ref{genfunction}) with respect to $s$ at $s=0$ we can easily find all the expectation values for arbitrary powers of the area:
$$
E_*[A^n\,|\,p]=(-1)^n\frac{\displaystyle[\nu^p]\left(\left.\frac{\partial^n}{\partial s^n}\right|_{s=0}G_*(\nu,s)\right)}{[\nu^p]G_*(\nu,0)}\,.
$$
In particular, the mean  and the variance
$$
\mu_{*p}=E_*[A\,|\,p]\,,\quad \sigma_{*p}^2=E_*[A^2\,|\,p]-E^2_*[A\,|\,p]
$$
of the area distribution conditioned by $P=p$, can be obtained in a straightforward way. Exact expressions (as closed functions of $p$) for $\mu_{*p}\,$,$\sigma^2_{*p}\,$, and the normalization factor
$$
\alpha_{*p}:=[\nu^p]G_*(\nu,0)=\sum_{c\in\mathcal{P}_p}d_*(c)
$$
can be found in Appendix \ref{App:A}. In the asymptotic regime $p\rightarrow \infty$ these objects follow very simple laws:
\begin{eqnarray}
\alpha_{*p}&\sim& \frac{2\nu_0^2}{(10\nu_0^2+3)}\frac{1}{\nu_0^p}\,,\\
\mu_{*p}&\sim& \mu \cdot p = (0.34959022\cdots)\cdot p\,,\label{mean*}\\
\sigma_{*p}^2&\sim& \sigma^2 \cdot p = (0.00009817\cdots) \cdot p\,,\label{variance*}
\end{eqnarray}
where $\nu_0=(0.77039825\cdots)$ will denote the single real root the polynomial $2\nu^5+\nu^3-1$. A closed expression for $\nu_0$ in terms of hypergeometric functions is given in \cite{Agullo:2010zz} (we will discuss some details concerning this issue in Section \ref{sect:substructure}). As we show in Appendix \ref{App:A}, the coefficients $\mu$ and $\sigma^2$ appearing in (\ref{mean*}) and (\ref{variance*}) can be written in terms of $\nu_0$. The linear growth of the variance and the fact that the spacing between successive steps tends to a constant value strongly suggests that the steps will fade as the area increases. This will be shown in detail in Section \ref{gaussian}.

We want to mention here that the leading behavior of $\mu_{*p}\sim (0.34959022\cdots)\cdot p$ \textit{exactly} coincides with the one obtained in \cite{Agullo:2010zz} by using a completely different approach relying on a continuum approximation. We will provide an alternative proof of these asymptotic results in  Section \ref{gaussian}.

\subsection{Moment-generating function: The Domagala-Lewandowski approach}\label{DL}

When the projection constraint is considered, the starting point is the master generating function $G(\nu;z,x_1,\ldots,)$ defined in (\ref{master}). By substituting $x_i=e^{-s\sqrt{p_i}}$ we obtain
\begin{equation}
G(\nu,s;z):=G(\nu,z;e^{-s\sqrt{p_1}},e^{-s\sqrt{p_2}},\ldots)=\frac{1}{1-\sum_{k=1}^\infty \nu^{3k+2}(z^k+z^{-k})e^{-s\sqrt{k(k+2)}}}\,.
\label{generatingfunction}
\end{equation}
The function $G(\nu,s;z)$ satisfies
$$
[z^0][\nu^p]G(\nu,s;z)=\sum_{c\in\mathcal{P}_p}d(c)e^{-s A(c)}\,
$$
and, hence, we have the following expression for the expectation value
\begin{equation}
E[\exp(-s A)\,|\,p]=\frac{[z^0][\nu^p]G(\nu,s;z)}{[z^0][\nu^p]G(\nu,0;z)}\,.
\label{momentgf}
\end{equation}
By computing the derivatives of (\ref{generatingfunction}) with respect to $s$, at $s=0$, we can easily find all the expectation values for arbitrary powers of the area
$$
E[A^n\,|\,p]=(-1)^n\frac{\displaystyle[z^0][\nu^p]\left(\left.\frac{\partial^n}{\partial s^n}\right|_{s=0}G(\nu,s;z)\right)}{[z^0][\nu^p]G(\nu,0;z)}\,.
$$
In this case, the mean and the variance
$$
\mu_{p}=E[A\,|\,p]\,,\quad \sigma_{p}^2=E[A^2\,|\,p]-E^2[A\,|\,p]
$$
of the area distribution conditioned by $P=p$ can be obtained with some extra work due to the presence of the $z$-variable in $G(\nu,s;z)$. Expressions for $\mu_{p}\,$,$\sigma^2_{p}\,$, and the normalization factor
$$
\alpha_p:=[z^0][\nu^p]G(\nu,0;z)=\sum_{c\in\mathcal{P}_p}d(c)
$$
can be found in Appendix \ref{App:B}. In particular, it is possible to prove that $\alpha_p=0$ for all odd values of $p$ and hence only the even values of $P(c)=p$ have to be considered.  In this case, in the asymptotic regime $p=2q\rightarrow \infty$ we have
\begin{eqnarray}
\alpha_{2q}&\sim&\frac{1}{1+\nu_0^2}\sqrt{\frac{\nu_0(1-\nu_0^6)}{\pi (10\nu_0^2+3)}}\frac{1}{\sqrt{q}\nu_0^{2q}}\,, \\
\mu_{2q}&\sim& 2\mu \cdot q = 2\cdot (0.34959022\cdots)\cdot q\,,\label{meanDL}\\
\sigma_{2q}^2&\sim& 2\sigma^2 \cdot q = 2\cdot (0.00009817\cdots) \cdot q\,.\label{varianceDL}
\end{eqnarray}

\bigskip

The statistical treatment given in this section suggests two approximate models for the behavior of the black hole entropy as a function of the horizon area. In the first one the steps in the entropy are approximated by Heaviside step functions with jumps of magnitude $\alpha_{p}$ (or, respectively $\alpha_{*p}$ when the projection constraint is ignored) located at areas given by the mean values $\mu_{p}$. The second, improved, model will use smoothed steps given by the (integrated) Gaussian distributions of mean $\mu_{p}$ and variance $\sigma_{p}^2$ with height $\alpha_{p}$. We discuss them in the following sections.

\section{Using the mean: The staircase approximation for the entropy}\label{staircase}

\begin{center}
\begin{figure}[htbp]
\hspace*{-1cm}
\includegraphics[width=18.5cm]{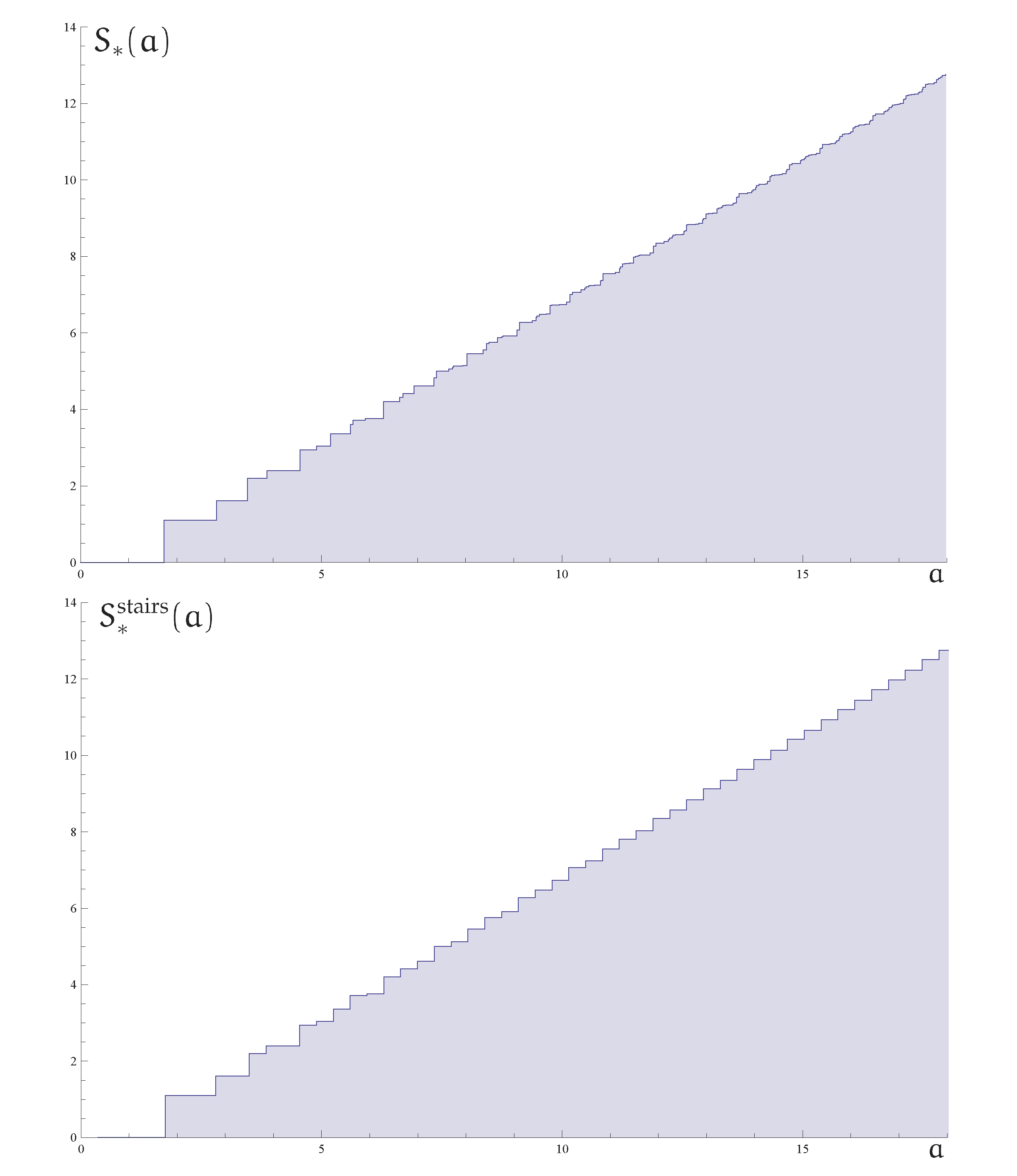}
\caption{The upper part of the figure shows the exact value of the entropy $S_*(a)$ for areas below $18$ (in units of $4\pi\gamma\ell_P^2$) in the case in which the projection constraint is ignored. The lower part represents the staircase approximation to the entropy given by $S^{\textrm{stairs}}_*(a)=\log\left(\sum_{p=0}^\infty \alpha_{p}\theta(a-\mu\cdot p)\right)\,.$} \label{Fig:esc_mean}
\end{figure}
\end{center}

A coarse approximation for the exponentiated entropy $\exp S_{*}(a)$ and $\exp S(a)$ can be obtained by assigning the sum of all the degeneracies corresponding to each peak to a single step located at the mean area value. This can be done by employing Heaviside step functions (denoted by $\theta$ in the following) in several slightly different ways (obtained by using the asymptotic approximations for $\mu_{*p}$, $\mu_{p}$, $\alpha_{*p}$ and $\alpha_{p}$):
\begin{eqnarray}
\begin{array}{lll}
\displaystyle\sum_{p=0}^\infty \alpha_{*p}\theta(a-\mu_{*p})\,,&\quad\quad\quad&\displaystyle\sum_{p=0}^\infty \alpha_{p}\theta(a-\mu_{p})\,,
\vspace*{2mm}\\
\displaystyle\sum_{p=0}^\infty \alpha_{*p}\theta(a-\mu p)\,,&&\displaystyle\sum_{p=0}^\infty \alpha_{p}\theta(a-\mu p)\,,\\
\displaystyle 1+\frac{2\nu_0^2}{(10\nu_0^2+3)}\sum_{p=5}^\infty \frac{1}{\nu_0^p}\theta(a-\mu p)\,,&&\displaystyle
1+\frac{1}{1+\nu_0^2}\sqrt{\frac{2\nu_0(1-\nu_0^6)}{\pi (10\nu_0^2+3)}}\sum_{p \,\textrm{ even}}^\infty \frac{1}{\sqrt{p}\nu_0^{p}}\theta(a-\mu p)\,,
\end{array}
\nonumber
\end{eqnarray}
where the first column corresponds to the case without the projection constraint. The validity of these approximations for small areas is clearly seen in Fig. \ref{Fig:esc_mean}.

By proceeding in this way the entropy will obviously display a staircase structure because we are approximating it as a sum of sharp, (asymptotically) equally spaced, steps. This means that this simplified approach will not be suitable to address the persistence (or lack thereof) of the structure seen in the entropy for small areas in the asymptotic regime. However, it can be used to estimate the value of the Immirzi parameter as $\gamma\approx \gamma_{\textrm{stairs}}$ because it provides a simple expression for the growth of the entropy as a function of the area. In fact, in the case where the projection constraint is neglected, we easily find
\begin{equation}
\pi\gamma_{\textrm{stairs}} := \frac{-\log \nu_0}{\mu }=(0.74615268\cdots).\label{gamma_stairs}
\end{equation}
This value must be compared with the one obtained by Meissner in \cite{Meissner:2004ju}
\begin{equation}
\pi\gamma=(0.74623179\cdots).
\label{gamma_meissner}
\end{equation}
Notice that $\gamma_{\textrm{stairs}}$ has nothing to do with the value of the  Immirzi parameter $\pi \gamma_{\textrm{flux}}=\log 3$ derived in the context of LQG from the  evenly spaced flux area operator used in \cite{FernandoBarbero:2009ai} which can be interpreted in terms of the Schwarzschild quasinormal modes \cite{Dreyer:2002vy,FernandoBarbero:2009ai}. Though one could argue that the value $\gamma_{\textrm{stairs}}$ derived for the Immirzi parameter in this approximation is quite good, the fact that it predicts a growing behavior, different from the true one, means that the entropy and its staircase approximation will diverge linearly.

A convenient way to derive (\ref{gamma_stairs}) is by using Laplace transform techniques. Let us discuss, in the first place, the staircase approximation without the projection constraint. To this end we consider
$$
\exp S^{\textrm{stairs}}_*(a):=1+\frac{2\nu_0^2}{(10\nu_0^2+3)}\sum_{p=5}^\infty \frac{1}{\nu_0^p}\theta(a-\mu p)
$$
whose Laplace transform $\mathcal{L}(\exp S^{\textrm{stairs}}_*,s)$  can be computed in closed form (as a function of the complex variable $s$)
$$
\mathcal{L}(\exp S^{\textrm{stairs}}_*,s)=\frac{1}{s}+\frac{2\nu_0^2}{(10\nu_0^2+3)s}\sum_{p=5}^\infty \frac{e^{-\mu p s}}{\nu_0^p}= \frac{1}{s}+ \frac{2}{(10\nu_0^2+3)\nu_0 ^2}  \frac{e^{-4 \mu s}}{ (\nu_0e^{\mu s}-1 )s}\,.
$$
The pole $s_{\textrm{stairs}}=-\log( \nu_0)/\mu $ is responsible for the exponential growth of $\exp S^{\textrm{stairs}}_*(a)$, in the regime $a\rightarrow \infty$, given by (\ref{gamma_stairs}). Notice that, in addition to this real pole (and $s=0$), there are infinitely many others of the form $-\log( \nu_0)/\mu +2k\pi i/\mu $, $k\in\mathbb{Z}\setminus\{0\}$ that account for the steps in this approximation for the entropy \cite{G.:2008mj}.

When the projection constraint is taken into account the configurations $c$ with odd values of $P(c)=p$ have zero degeneracy and hence only even values of $p=2q$ have to be considered. The staircase approximation is then
$$
\exp S^{\textrm{stairs}}(a):=1+\frac{1}{1+\nu_0^2}\sqrt{\frac{2\nu_0(1-\nu_0^6)}{\pi (10\nu_0^2+3)}}\sum_{q=1}^\infty \frac{1}{\sqrt{2q}\nu_0^{2q}}\theta(a-2\mu q)\,.
$$
The Laplace transform $\mathcal{L}(\exp S^{\textrm{stairs}},s)$ is given by
$$
\mathcal{L}(\exp S^{\textrm{stairs}},s)=\frac{1}{s}+\frac{1}{1+\nu_0^2}\sqrt{\frac{\nu_0(1-\nu_0^6)}{\pi (10\nu_0^2+3)}}\mathrm{Li}_{\textrm{\textonehalf}}(e^{-2(\mu s+\log \nu_0)})
$$
where $\mathrm{Li}_{\textrm{\textonehalf}}$ denotes the polylogarithm of order \textonehalf. The singularities of $\mathrm{Li}_{\textrm{\textonehalf}}(e^{-2(\mu s+\log \nu_0)})$ are branch cuts starting at the \textit{same} straight line  $\textrm{Re}(s)=s_{\textrm{stairs}}$ in the  complex $s$-plane as the singularities found for the case without the projection constraint: $-\log( \nu_0)/\mu +k\pi i/\mu $, $k\in\mathbb{Z}$. Notice that the spacing between these points is half the one obtained when the projection constraint is ignored. This means that the width of the steps doubles in this case. The effect of the branch cuts is to modify the asymptotic behavior of the entropy by the addition of the expected logarithmic corrections, however, the linear growth is the same as before and the inferred value of the Immirzi parameter is still given by (\ref{gamma_stairs}).

The failure to reproduce the exact value for the Immirzi parameter in this approximation stems from the fact that, for a given value of the area $a$, the model neglects to take into account contributions coming from peaks with $p$ beyond the largest one satisfying $\mu p\leq a$. It also misses some contributions coming from lower values of $p$ (at least in the asymptotic regime of large areas).

\section{Using the mean and the variance: Smoothed gaussian approximation for the entropy}\label{gaussian}

An improved model for the black hole entropy can be obtained by approximating the steps by Gaussian distributions with mean and variance given by (\ref{mean*}) and (\ref{variance*}). This will take into account the fact that the steps become wider with increasing values of $p$ (an effect that can be readily seen by plotting the exact values of the entropy for small black holes as functions of the area). This is obviously relevant to study whether the staircase structure is present in the asymptotic limit. At this point it is just appropriate to quote from page 611 of the book by Flajolet and Sedgewick \cite{Flajolet}

\bigskip

\textit{``Many applications, in various sciences as well as in combinatorics itself, require
quantifying the behaviour of parameters of combinatorial structures. The corresponding
problems are now of a multivariate nature, as one typically wants a way to estimate
the number of objects in a combinatorial class having a fixed size and a given parameter
value. Average-case analyses usually do not suffice, since it is often important to
predict what is likely to be observed in simulations or on actual data that obey a given
randomness model, in terms of possible deviations from the mean —-this signifies that
information on probability distributions is wanted. [...] Indeed, it is frequently observed that the histograms of
the distribution of a combinatorial parameter (for varying size values) exhibit a common
characteristic ``shape'', as the size of the random combinatorial structure tends
to infinity. In this case, we say that there exists a limit law.''}

\bigskip

In our case we have a multivariate combinatorial problem where both the area and the peak parameter $P$ play a significant role. Furthermore, we have that the distribution of one of the parameters (the area of the peaks) displays a characteristic shape as the peak counter grows towards infinity. As we will show in this section the methods appearing in \cite{Flajolet} will allow us to gather important information about the behavior of the entropy as a function of the area. In particular, we will see that a Gaussian law --reminiscent of the Central Limit Theorem of probability theory-- plays an important role in the analysis presented here.

\subsection{Gaussian law for the peaks}

\begin{center}
\begin{figure}[htbp]
\includegraphics[width=16.5cm]{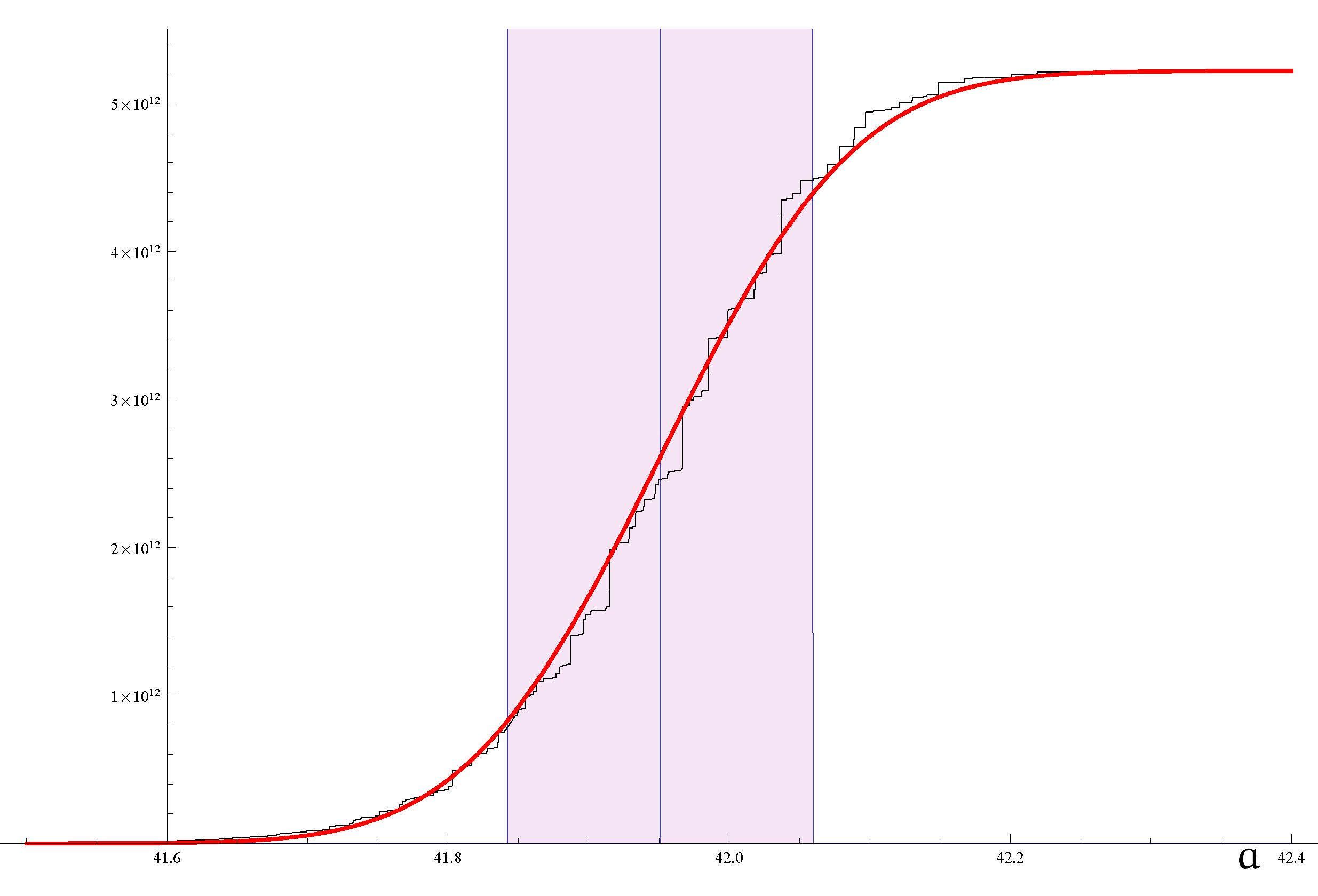}
\caption{Comparison between the exact shape of the step $\sum_{a'\leq a} D_*(a'\,|\,120)$ and the corresponding smoothed Gaussian approximation $\frac{\alpha_{*120}}{2}\left(1+\mathrm{erf}\left(\frac{a-\mu_{*120}}{\sqrt{2}\sigma_{*120}}\right)\right)\,
$. The vertical band is centered around the value $\mu_{*120}$ and has a width of $2\sigma_{*120}$.} \label{Fig:peak_gauss}
\end{figure}
\end{center}

The key idea --in the case where the projection constraint is neglected\footnote{The case when the projection constraint is taken into account can be handled by adapting theorem IX.12 of \cite{Flajolet}.}-- is to use theorem IX.9 (page 656) of \cite{Flajolet} for the generating function $G_*(\nu,s)$ given in (\ref{genfunction}). The theorem tells us that the mean and the variance for $P=p$ can be easily obtained in terms of the ``analytic'' mean\footnote{Notice that the minus sign in our definition of $\mathfrak{m}(f_*)$ originates in our sign convention for the variable $s$ appearing in our moment-generating functions.} $\mathfrak{m}(f_*)$ and variance $\mathfrak{v}(f_*)$ of a function $f_*$ as
\begin{eqnarray}
\mu_{*p}&=&\mathfrak{m}(f_*) p +O(1)=-\frac{f'_*(0)}{f_*(0)}\cdot p+O(1)\,,\label{media2}\\
\sigma_{*p}^2&=&\mathfrak{v}(f_*)p+O(1)=\left(\frac{f''_*(0)}{f_*(0)}-\left(\frac{f'_*(0)}{f_*(0)}\right)^2\right)p + O(1)\,.\label{varianza2}
\end{eqnarray}
The function
$$
f_*(s):=\frac{\nu_*(0)}{\nu_*(s)}\,, \quad  f_*(0)=1\,,
$$
is given in terms of  $\nu_*(s)$ defined by $\nu_*(0)=\nu_0$ and $Q_*(\nu_*(s),s)=0$, where
$$
Q_*(\nu,s):=\frac{1}{G_*(\nu,s)}=1-2\sum_{k=1}^\infty \nu^{3k+2}\exp(-s\sqrt{k(k+2)})\,.
$$
The implicit function theorem allows us to obtain a power series expansion in terms of the variable $s$ with coefficients given by derivatives of $Q_*(\nu,s)$ evaluated at $\nu=\nu_0$ and $s=0$. Explicitly
\begin{eqnarray*}
\nu_*(s)&=&\nu_0-\frac{q_{* 0,1}}{q_{* 1,0}} s-\frac{q_{* 1,0}^2q_{* 0,2}-2q_{* 1,0}q_{* 1,1}q_{* 0,1}+q_{* 2,0}q_{* 0,1}^2}{2q_{* 1,0}^3} s^2+O(s^3)\\
\end{eqnarray*}
with
$$
q_{* i,j}:=\left.\frac{\partial^{i+j}Q_*}{\partial \nu^i\partial s^j}\right|_{(\nu_0,0)}\,.
$$
The results given by (\ref{media2}) and (\ref{varianza2}) are  \textit{exactly} the same that we have found above in equations (\ref{mean*}) and (\ref{variance*}). In any case this is a very efficient method to compute the numerical values of the mean and the variance of the peak distributions. The theorem, however, provides us with another very important convergence result: The random variable
$$X_p:=\frac{A_p-\mu_{*p}}{\sigma_{*p}}$$
with (normalized) distribution function
$$
F_{*p}(x)=\mathrm{Prob}_{*}(X_p\leq x)=\frac{\sum_{c\in \mathcal{P}_p\cap X^{-1}_p((-\infty,x])} d_*(c)}{\sum_{c\in \mathcal{P}_p} d_*(c)}=\frac{\sum_{x'\leq x}\sum_{c\in \mathcal{P}_p\cap X^{-1}_p(x')} d_*(c)}{\alpha_{*p}}
$$
converges, pointwise, to a Gaussian distribution
$$
\lim_{p\rightarrow\infty}F_{*p}(x)=\frac{1}{2}\left(1+\mathrm{erf}\left(\frac{x}{\sqrt{2}}\right)\right)
:=\frac{1}{2}\left(1+\frac{2}{\sqrt{\pi}}\int_{0}^{x/\sqrt{2}}e^{-t^2}\mathrm{d}t \right)\,,
$$
with a $O(1/\sqrt{p})$ speed of convergence. In terms of the area this fact implies that we can write
$$
\sum_{a'\leq a} D_*(a'\,|\,p)=\sum_{a'\leq a}\sum_{c\in \mathcal{C}_p(a')} d_{*}(c)=\sum_{c\in \mathcal{P}_p\cap A^{-1}_p(0,a]}d_*(c)= \alpha_{*p} F_{*p}\left(\frac{a-\mu_{*p}}{\sigma_{*p}}\right)\,.
$$
In practice this tells us that each smoothed step, given by the function
$$
a\mapsto \frac{\alpha_{*p}}{2}\left(1+\mathrm{erf}\left(\frac{a-\mu_{*p}}{\sqrt{2}\sigma_{*p}}\right)\right)\,,
$$
is a good approximation (see Fig. \ref{Fig:peak_gauss}) to the actual shape of the graph of the function
$$a\mapsto \displaystyle \sum_{a'\leq a} D_*(a'\,|\,p)$$
appearing in the definition of the entropy (\ref{D}). This approximation improves as $p$ grows.

If one compares, instead, the graphs of the functions
$$
a\mapsto \frac{\alpha_{*p}}{\sqrt{2\pi}\sigma_{*p}}\exp\left(-\frac{(a-\mu_{*p})^2}{2\sigma_{*p}^2}\right)\quad \textrm{and}\quad
a\mapsto \displaystyle D_*(a\,|\,p)\,,$$
corresponding to the peaks in the degeneracy spectrum, the Gaussian shape does not correspond in any way to an ``envelope'' of the actual peak defined by the degeneracies (see Fig. \ref{Fig:intro}), although the maxima appear roughly for the same value of the area and the widths match reasonably. We want to point out that the parameters of the Gaussian approximation have been obtained \textit{a priori} from the moment-generating function. So we are not fitting the ``peak data'' to a Gaussian but, rather, deriving the statistical properties of the distribution that they define by relying on an exact statistical analysis.

\subsection{Gaussian approximation for the entropy}

\begin{center}
\begin{figure}[htbp]
\hspace*{-1cm}
\includegraphics[width=18.5cm]{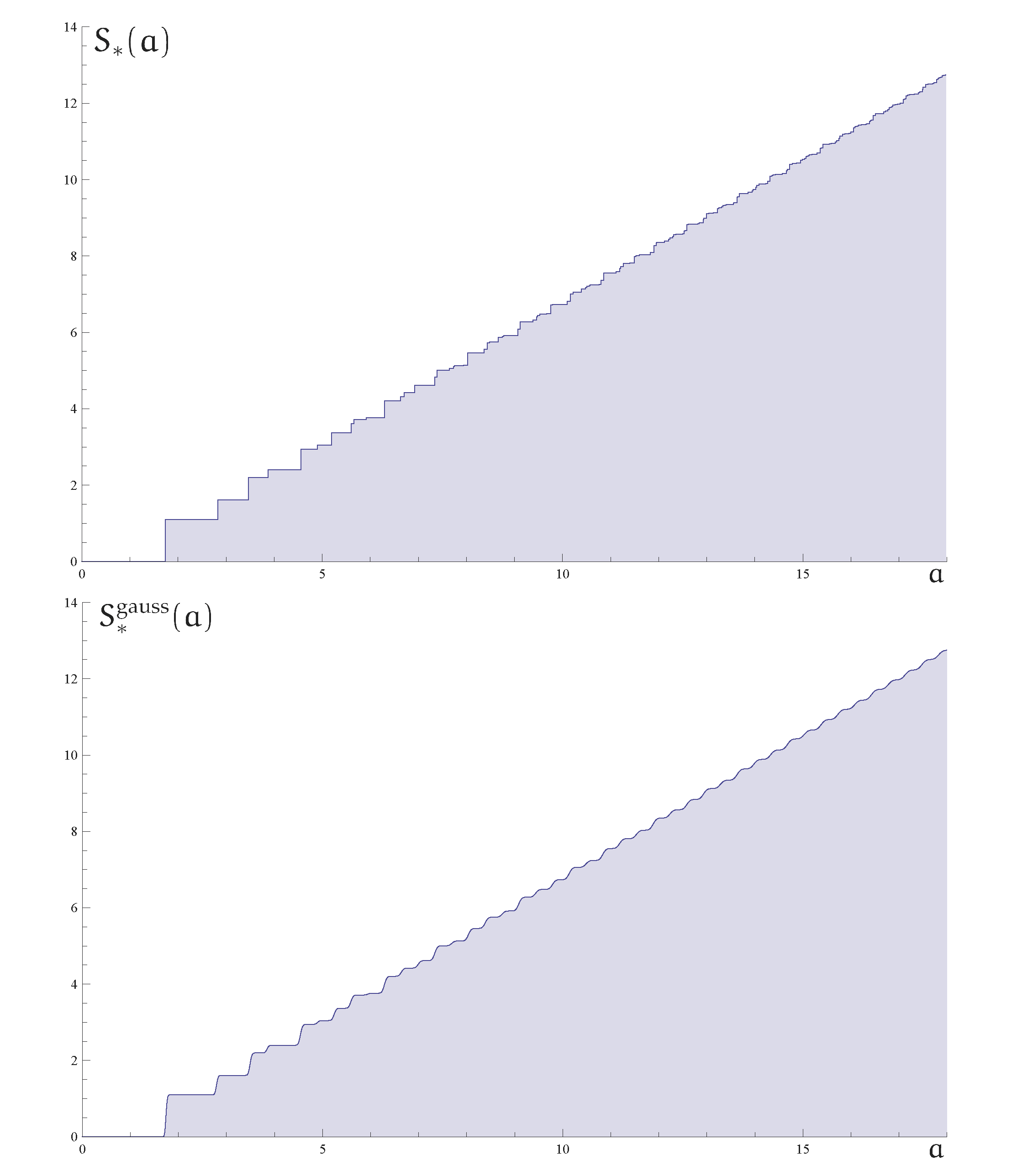}
\caption{The upper part of the figure shows the exact value of the entropy $S_*(a)$ for areas below $18$ (in units of $4\pi\gamma\ell_P^2$) in the case in which the projection constraint is ignored. The lower part represents the Gaussian approximation $S^{\textrm{gauss}}_*(a)=\log\left(1+\sum_{p=1}^{\infty}\frac{\alpha_{*p}}{2}
\left(1+\mathrm{erf}\left(\frac{a-\mu p}{\sqrt{2p}\sigma }\right)\right)\right)\,.$} \label{Fig:esc_gauss_1}
\end{figure}
\end{center}

\begin{center}
\begin{figure}[htbp]
\hspace*{-1cm}
\includegraphics[width=18.5cm]{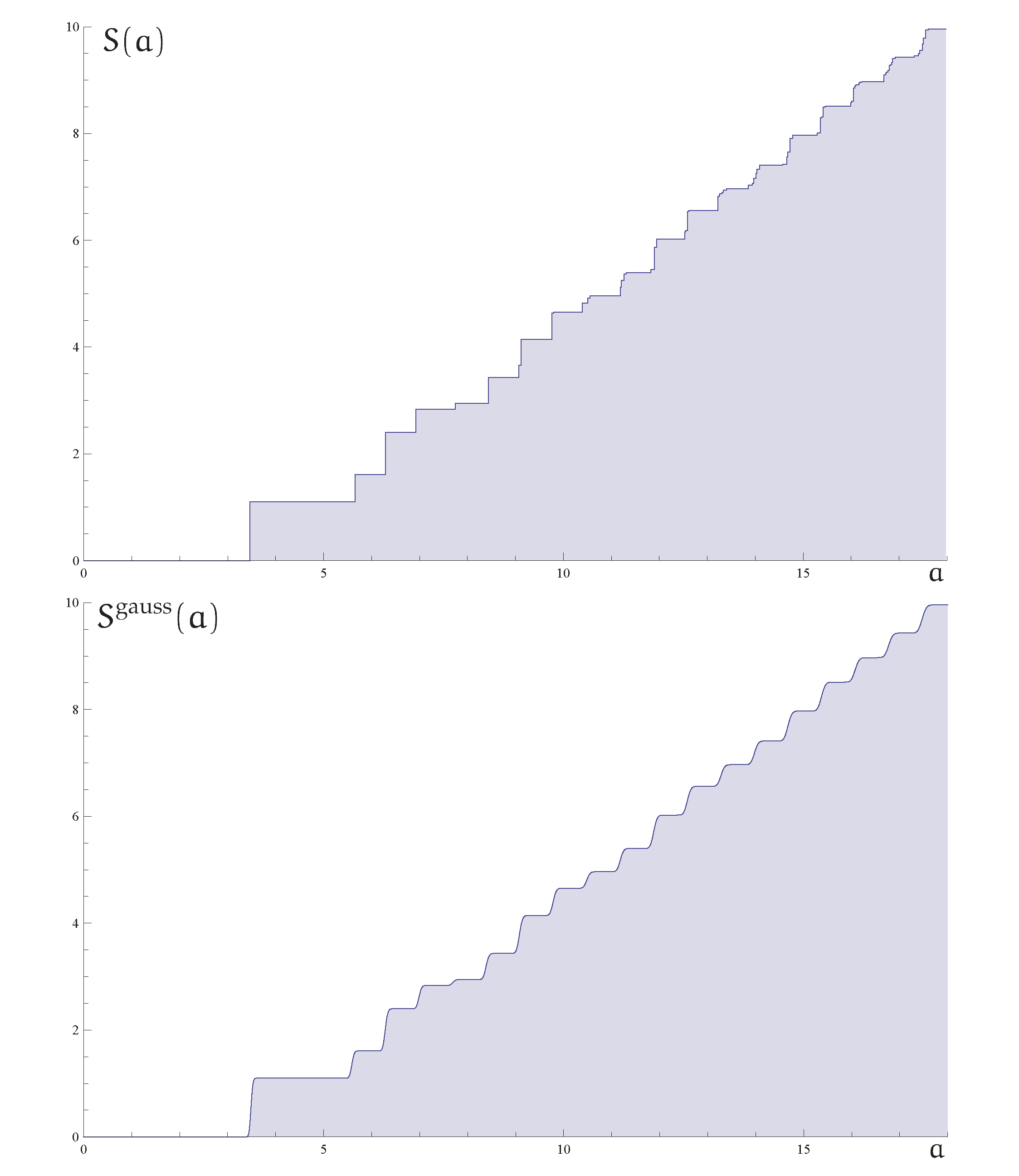}
\caption{The upper part of the figure shows the exact value of the entropy $S(a)$ for areas below $18$ (in units of $4\pi\gamma\ell_P^2$) in the case in which the projection constraint is taken into account. The lower part represents the Gaussian approximation $S^{\textrm{gauss}}(a)=\log\left(1+\sum_{p=1}^{\infty}\frac{\alpha_{p}}{2}
\left(1+\mathrm{erf}\left(\frac{a-\mu p}{\sqrt{2p}\sigma }\right)\right)\right)\,.$ The staircase structure decays more slowly now as discussed in the text.} \label{Fig:esc_gauss_2}
\end{figure}
\end{center}

The idea now is to model the entropy as a sum of smoothed steps like the ones shown in Figure \ref{Fig:peak_gauss}. However, it is very important to be aware of the fact that the convergence of the individual steps to their Gaussian approximations does not guarantee the convergence of their sum to the actual value of the entropy. Let us consider then approximations for the exponentiated entropy $\exp{S_*(a)}$ and $\exp{S(a)}$ obtained by adding Gaussian steps. In the case where the projection constraint is ignored these are
\begin{eqnarray*}
&&1+\sum_{p=1}^{\infty}\frac{\alpha_{*p}}{2}
\left(1+\mathrm{erf}\left(\frac{a-\mu_{*p}}{\sqrt{2}\sigma_{*p}}\right)\right)\,,\\
&&1+\sum_{p=1}^{\infty}\frac{\alpha_{*p}}{2}
\left(1+\mathrm{erf}\left(\frac{a-\mu p}{\sqrt{2p}\sigma }\right)\right)\,,\\
&& 1+\frac{\nu_0^2}{(10\nu_0^2+3)}\sum_{p=5}^{\infty} \frac{1}{\nu_0^p}\left(1+\mathrm{erf}\left(\frac{a-\mu p}{\sqrt{2p}\sigma }\right)\right)\,.
\end{eqnarray*}
Similar expressions (to be discussed later) hold for the case in which the projection constraint is incorporated. Figures \ref{Fig:esc_gauss_1} and \ref{Fig:esc_gauss_2} show a comparison between these approximations and the actual value of the entropy for the smallest areas. As it can be seen, the agreement is excellent. Not only the height of the steps is reproduced with high fidelity but also their progressive smoothing. In the case when the projection constraint is taken into account the staircase structure is more evident.

In spite of this remarkable agreement we know, as we have learned in the preceding section, that the asymptotic growth of the Gaussian approximation may not be the exact one (i.e. the value of $\gamma_{\textrm{gauss}}$ derived here may differ from the true one). In fact, this will be shown to be the case. In order to study the asymptotic growth of the Gaussian approximation to the entropy it suffices to consider
$$
\exp S^{\textrm{gauss}}_*(a):=1+\frac{\nu_0^2}{(10\nu_0^2+3)}\sum_{p=5}^{\infty} \frac{1}{\nu_0^p}\left(1+\mathrm{erf}\left(\frac{a-\mu p}{\sqrt{2p}\sigma }\right)\right)\,,
$$
and study the singularity structure of its Laplace transform written in terms of the complex variable $s$. This is given by
\begin{eqnarray}
\mathcal{L}(\exp S^{\textrm{gauss}}_*,s)&=&\frac{1}{s}
+\frac{\nu_0^2}{(10\nu_0^2+3)s}\sum_{p=5}^{\infty} \frac{1}{\nu_0^p}\left(1-\mathrm{erf}\left(\frac{\mu}{\sqrt{2}\sigma }\sqrt{p}\right)\right)\label{laplacegaussian}\\
&+&\frac{\nu_0^2}{(10\nu_0^2+3)s}\sum_{p=5}^{\infty} \frac{1}{\nu_0^p}
\left(1+\mathrm{erf}\left(\frac{\sigma^2s-\mu}{\sqrt{2}\sigma }\sqrt{p}\right)\right)
\exp\left(\Big(\frac{\sigma^2}{2}s^2-\mu s\Big)p\right)\nonumber\,.
\end{eqnarray}
Notice that despite the fact that the Laplace transforms of the individual steps are entire functions in the complex variable $s$, the analytic extension of their sum (restricted to the values of $s$ for which it actually converges) may, and actually will, have singularities. By looking at the inversion formula for Laplace transforms it is easy to see that these singularities determine the asymptotic behavior of the original sum. This is discussed in detail in Appendix \ref{App:C}. The first two terms in (\ref{laplacegaussian}) have a very simple analytic structure because they just have a pole at $s=0$.
The second sum in (\ref{laplacegaussian}) is more complicated and may converge or diverge depending on the values of $s$. It is easy to see that it converges for all $s\in \mathbb{C}$ such that $|\mathrm{arg}(s)|\leq 3\pi/4$ (the range of the argument is taken to be $\mathrm{arg}(s)\in(-\pi,\pi]$). Inside the wedge $|\mathrm{arg}(s)|>3\pi/4$ the series converges  for values of $s$ to the right of the hyperbola (see Figure \ref{Fig:sing} in Appendix \ref{App:C})
\begin{equation}
\mathrm{Re}\Big(\frac{\sigma^2}{2}s^2-\mu s-\log \nu_0\Big)=0\,.\label{hyperbola}
\end{equation}
This divergence (in the region  $|\mathrm{arg}(s)|>3\pi/4$) is due to a term of the form
\begin{eqnarray}
&&\frac{2\nu_0^2}{(10\nu_0^2+3)s}\sum_{p=5}^{\infty} \frac{1}{\nu_0^p}
\exp\left(\Big(\frac{\sigma^2}{2}s^2-\mu s\Big)p\right)\,.
\label{divergente}
\end{eqnarray}
This means that by subtracting this expression from the series that we are looking at, we get another series that converges in the full wedge $|\mathrm{arg}(s)|>3\pi/4$ to a function $h(s)$ (with no singularities). The sum (\ref{divergente}) can actually be performed in closed form to get a meromorphic extension to  $|\mathrm{arg}(s)|>3\pi/4$ of the function that it defines inside its region of convergence. This is given by
\begin{equation}
\frac{\displaystyle 2\exp\left(\frac{5\sigma^2}{2}s^2-5\mu s\right)}
{\displaystyle \nu^3_0(10\nu_0^2+3)\left(1-\exp\Big(\frac{\sigma^2}{2}s^2-\mu s-\log\nu_0\Big)\right)s}\,.
\label{extension}
\end{equation}
The analytic extension of the Laplace transform (\ref{laplacegaussian}) to the wedge is then given by the sum of the first two terms in (\ref{laplacegaussian}), the function $h(s)$ and (\ref{extension}). Hence the singularities of the Laplace transform (\ref{laplacegaussian}) are $s=0$ and those of (\ref{extension}). These are isolated simple poles located on the hyperbola (\ref{hyperbola})  defined above (see Fig. \ref{Fig:sing}) and given by the condition
$$
\frac{\sigma^2}{2}s^2-\mu s-\log\nu_0=2k\pi i\,,\quad k\in\mathbb{Z}\,.
$$
The single real pole at
$$
s_{\textrm{gauss}}=\frac{\mu }{\sigma^2}\left(1-\sqrt{1+\frac{2\sigma^2}{\mu ^2}\log\nu_0}\right)
$$
dictates the asymptotic growth of the entropy in this Gaussian approximation and gives an improved estimate of the Immirzi parameter
\begin{equation}
\pi\gamma_{\textrm{gauss}}=s_{\textrm{gauss}}=(0.74623087\cdots)\label{gamma_gauss}
\end{equation}
to be compared with the actual value $\pi\gamma=(0.74623179\cdots)$. As it can be seen, the value of $\gamma_{\textrm{gauss}}$ is better than $\gamma_{\textrm{stairs}}$  obtained in the preceding section by using the staircase approximation  but still not the true one $\gamma$, as they differ starting at the sixth decimal figure.

The decay of the staircase structure is dictated by the two poles with the smallest non-zero imaginary parts. These are given by
$$
s_{*\pm}:=\frac{\mu}{\sigma^2}\left(1-\sqrt{1+\frac{2\sigma^2}{\mu^2}(\log\nu_0\pm 2\pi i)}\right)=(0.70084660\cdots)\mp(17.97653845\cdots) i.
$$

\begin{center}
\begin{figure}[htbp]
\hspace*{-1cm}
\includegraphics[width=18.5cm]{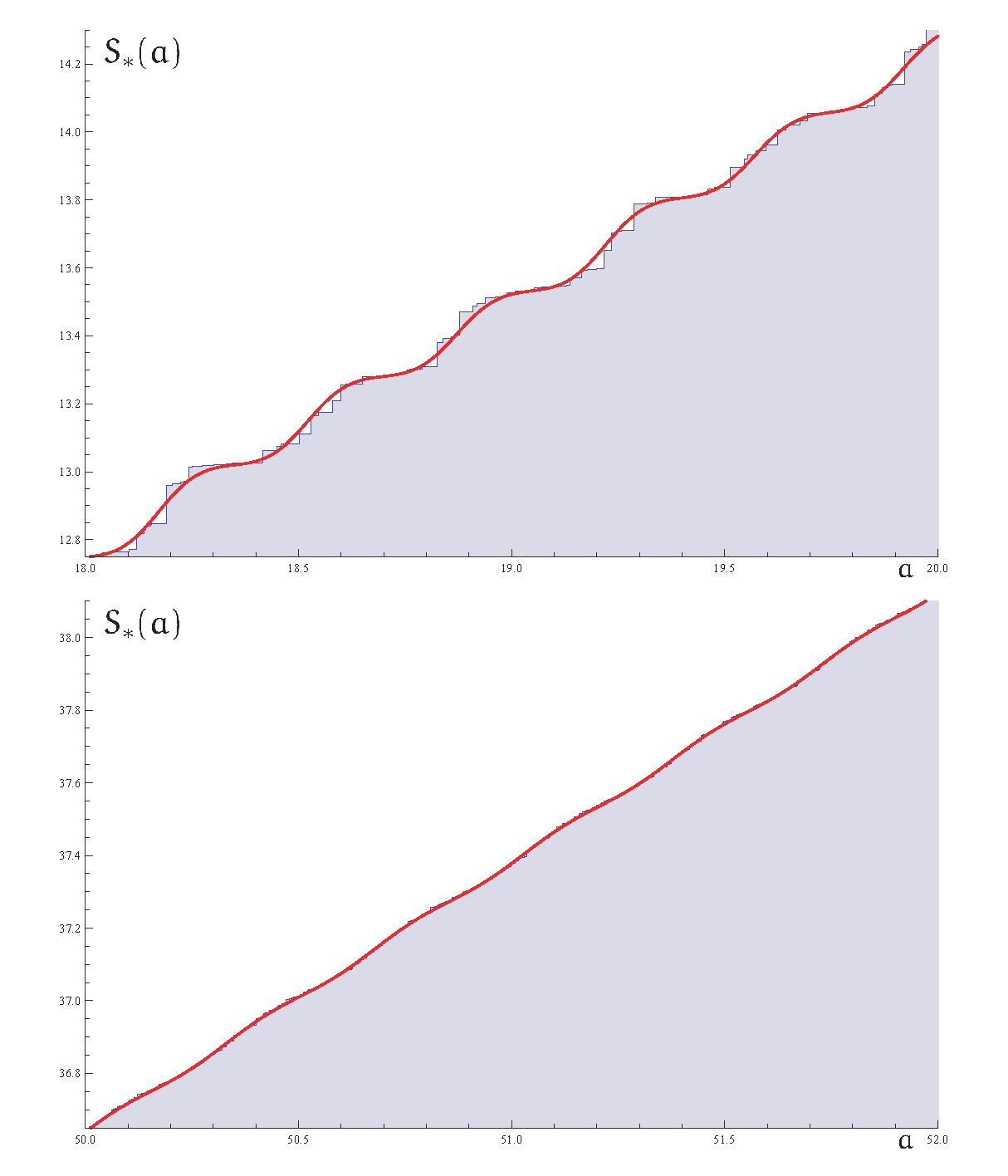}
\caption{Comparison between the exact values of the entropy $S_*$ and their Gaussian approximation for two different area intervals. The necessary computations are made possible by the introduction of the master generating functions (\ref{master*}). We can see that the Gaussian approximation works remarkably well in both cases. The fading of the oscillations in the entropy for the largest area interval is evident. Similar plots for area intervals around 70 show an essentially linear growth. } \label{Fig:esc_gauss_3}
\end{figure}
\end{center}

The magnitude of the imaginary part in the previous expression is very close to $2\pi/\mu=(17.97300040\cdots)$ as expected. Finally the comparison between $s_{\textrm{gauss}}$ and the real part of $s_{*\pm}$ tells us the decay rate of the staircase structure of the entropy. This given, essentially, by
\begin{equation}
\exp\Big(-\big(s_{\textrm{gauss}}-\mathrm{Re}(s_{*\pm})\big)a\Big)=\exp\Big(-(0.04538426\cdots)a\Big)\,
\label{decaystar}
\end{equation}
which means that the steps should have faded significantly for areas with an order of magnitude given by $1/(0.04538426\cdots)\sim 25$. We show in Fig. \ref{Fig:esc_gauss_3} the exact behavior of the entropy for two different area intervals in the case when the projection constraint is ignored and the corresponding Gaussian approximations. It can be readily seen both the accuracy of the Gaussian approximation in this regime and the decay of the staircase structure (essentially absent for areas around $50$). Another interesting feature that can be seen in Fig. \ref{Fig:esc_gauss_3} is the fact that as the density of the area spectrum increases, the jumps in the values of the entropy for consecutive area eigenvalues become smaller and smaller. Hence the entropy is better and better described by a smooth curve (in fact a straight line). It should be pointed out that the BH degeneracy spectrum at this regime still shows a distinct peak structure produced by configurations of very large degeneracy that, however, give an almost negligible contribution to the total degeneracy of the individual peaks for large areas.
When the projection constraint is taken into account the Gaussian approximation is
$$
\exp S^{\textrm{gauss}}(a):=1+\frac{1}{2(1+\nu_0^2)}\sqrt{\frac{\nu_0(1-\nu_0^6)}{\pi (10\nu_0^2+3)}}\sum_{q=1}^{\infty} \frac{1}{\sqrt{q}\nu_0^{2q}}\left(1+\mathrm{erf}\left(\frac{a-2\mu q}{2\sqrt{q}\sigma }\right)\right)\,,
$$
and the singularities of its Laplace transform --that again control the asymptotic behavior of the entropy-- are encoded in the series
$$
\sum_{q=1}^{\infty} \frac{1}{\sqrt{q}\nu_0^{2q}}
\exp\left(2\Big(\frac{\sigma^2}{2}s^2-\mu s\Big)q\right)=\mathrm{Li}_{\textrm{\textonehalf}}\left(\exp\Big(\sigma^2s^2-2\mu s-2\log\nu_0\Big)\right)
$$
that plays the same role as (\ref{divergente}) when the projection constraint was ignored.
The singularities satisfy now
$$
\frac{\sigma^2}{2}s^2-\mu s-\log\nu_0=k\pi i\,,\quad k\in\mathbb{Z}\,.
$$
The single real branch point is located at $s_{\textrm{gauss}}$, which is precisely the place where the real pole in the case without the projection constraint is placed. The change of the singularity nature (branch points instead of poles) means that the asymptotic behavior of the entropy in the Gaussian approximation (with the projection constraint) will acquire the expected logarithmic corrections.  The decay of the staircase structure is controlled by the two singularities with the smallest non-zero imaginary parts. These are given by
$$
s_{\pm}:=\frac{\mu}{\sigma^2}\left(1-\sqrt{1+\frac{2\sigma^2}{\mu^2}(\log\nu_0\pm \pi i)}\right)=(0.73488453\cdots)\mp(8.98835516\cdots) i\,.
$$
Two important points must be mentioned now. First we notice that the real part of these singularities is much closer to $s_{\textrm{gauss}}$ than in the case without the projection constraint. The decay rate of the staircase structure is controlled now by
\begin{equation}
\exp\Big(-\big(s_{\textrm{gauss}}-\mathrm{Re}(s_{\pm})\big) a\Big)=\exp\Big(-(0.01134633\cdots)a\Big)\,.
\label{decay}
\end{equation}
This means that the staircase structure of the entropy will be perceptible for areas roughly four times as large as those corresponding to the non-projection constraint case. Hence, similar results to those shown in Fig. \ref{Fig:esc_gauss_3} are obtained with the projection constraint for areas around 200 in our units. Finally, the imaginary part of the singularities --that controls the width of the steps-- is essentially half the value that it had when the projection constraint was ignored and hence the steps are twice as wide.

\section{Other partitions}\label{sect:substructure}

An interesting question that naturally arises concerns the comparison of the results obtained with the peak counter $P=3K+2N$ that we have used throughout the paper with the ones obtained with other partitions defined by linear counters of the type $$P(\alpha,\beta):=\alpha K+\beta N\,,\quad \alpha,\,\beta\in \mathbb{N}.$$
An important point that we want to emphasize here is the fact that all these counters provide partitions of the space of black hole configurations, and hence the full black hole spectrum can be recovered by using any of them (and taking into account all its possible values). However, some of them may be better suited to understand or isolate specific features of the spectrum (such us the observed bands for low areas).

By reasoning as in \cite{Agullo:2008eg}, and relying on the solutions to the Pell equation, it is actually possible to find other counters, such as $P(5,4)=5K+4N$ or $P(7,5)=7K+5N$, that could be potentially useful to understand the degeneracy spectrum. In particular, we have seen that $P(3,2)$ works very well for small areas but leads to a Gaussian approximation that underestimates the value of the Immirzi parameter. It is then natural to wonder if a better counter could exist that provides a better estimate for $\gamma$ and still explains the low area behavior of the entropy. Even if the low area behavior is not captured by such a counter it could be used to understand the asymptotic limit of large areas. In this section we compare the $P(3,2)$ counter with $P(\alpha,\beta)$ and show that our original choice is the best one. This does not mean that these other counters are not useful themselves. In fact, we will show that they can also be used to refine the partition provided by $P(3,2)$ and study the interesting substructure of the peaks defined with the help of $P(3,2)$.

\subsection{Assessing the "goodness" of $P(3,2)=3K+2N$}

The level sets $\mathcal{P}_p(\alpha,\beta)=P(\alpha,\beta)^{-1}(p)$ of the function $P(\alpha,\beta)=\alpha K+\beta N$ lead to the partition $\mathcal{C}=\bigcup_p \mathcal{P}_p(\alpha,\beta)$ of the configuration space. It is important to point out that the partitions defined by $P(n\alpha,n\beta)$ are equivalent to the partition defined by $P(\alpha,\beta)$ for any positive integer $n$. This means that we can consider only values of $\alpha$ and $\beta$ that are coprime, i.e. such that $\gcd(\alpha,\beta)=1$.

By proceeding as in previous sections we arrive at the following peak generating function\footnote{For the sake of simplicity, we will work without the projection constraint here because the leading terms for the mean and the variance are not sensitive to it.}
\begin{eqnarray}
G_*(\nu,s;\alpha,\beta):=\frac{1}{1-2\sum_{k=1}^\infty \nu^{\alpha k+\beta}e^{-s\sqrt{k(k+2)}}}\,.
\label{generatingAlpha}
\end{eqnarray}
The mean and variance of the area distribution conditioned by $P(\alpha,\beta)=p$ can be easily derived from this generating function and have an asymptotic behavior given by
$$
\mu_p(\alpha,\beta)\sim \mu(\alpha,\beta)\cdot p\,,\quad \sigma^2_p(\alpha,\beta)\sim \sigma^2(\alpha,\beta)\cdot p\,,
$$
where the coefficients $\mu(\alpha,\beta)$ and $\sigma^2(\alpha,\beta)$ can be written now in terms of the the real root $\nu_0(\alpha,\beta)$ of the polynomial $2\nu^{\alpha+\beta}+\nu^{\alpha}-1$. For example, by following  \cite{Flajolet}, it is possible to write
\begin{eqnarray*}
\mu(\alpha,\beta) &=&\frac{4\nu_0^{\alpha+2\beta}(\alpha,\beta)}
                       {2\beta \nu_0^{\alpha+\beta}(\alpha,\beta)+\alpha}\sum_{k=1}^\infty \sqrt{k(k+2)}\nu_0^{\alpha k}(\alpha,\beta)
\end{eqnarray*}
and
\begin{eqnarray*}
\sigma^2(\alpha,\beta)&=&  \frac{\left(1-\nu_0^{\alpha }\right)^3  \left(\alpha ^2 \left(\nu_0^{\alpha }+1\right)+\beta^2
   \left(\nu_0^{\alpha }-1\right)^2-2 \alpha  \beta  \left(\nu_0^{\alpha }-1\right)\right)}{\nu_0^{2 (\alpha +\beta )}\left(\alpha
   +\beta-\beta  \nu_0^{\alpha } \right)^3} \left(\sum_{k=1}^\infty \sqrt{k(k+2)} \nu_0^{\alpha k+\beta} \right)^2\\
&-&\frac{2 \left(1-\nu_0^{\alpha }\right)^4 }{\nu_0^{2 (\alpha +\beta )}\left(\alpha +\beta-\beta  \nu_0^{\alpha } \right)^2} \left(\sum_{k=1}^\infty (\alpha k+\beta)\sqrt{k(k+2)} \nu_0^{\alpha k+\beta}\right) \left(\sum_{k=1}^\infty \sqrt{k(k+2)} \nu_0^{\alpha k+\beta} \right)
\\
&+& \frac{\nu_0^{\alpha }-3}{\left(\nu_0^{\alpha }-1\right) \left(\alpha +\beta-\beta  \nu_0^{\alpha } \right)}\,,
\end{eqnarray*}
where $\nu_0=\nu_0(\alpha,\beta)$. The root $\nu_0(\alpha,\beta)$ can be obtained as the value at $x=1$ of a suitable analytic extension of the function defined within its convergence disk by the Taylor series
$$
\sum_{n=0}^\infty \frac{2^n}{\beta n+1}\binom{-(\beta n+1)\alpha^{-1}}{n}x^{\beta n+1}=\sum_{n=0}^\infty \frac{(-2)^n}{\beta n+1}\binom{\big((\alpha+\beta)n+1-\alpha\big)\alpha^{-1}}{n}x^{\beta n+1}\,.
$$
These extensions are finite sums of hypergeometric functions.

\begin{center}
\begin{figure}[htbp]
\hspace*{0cm}
\includegraphics[width=16.5cm]{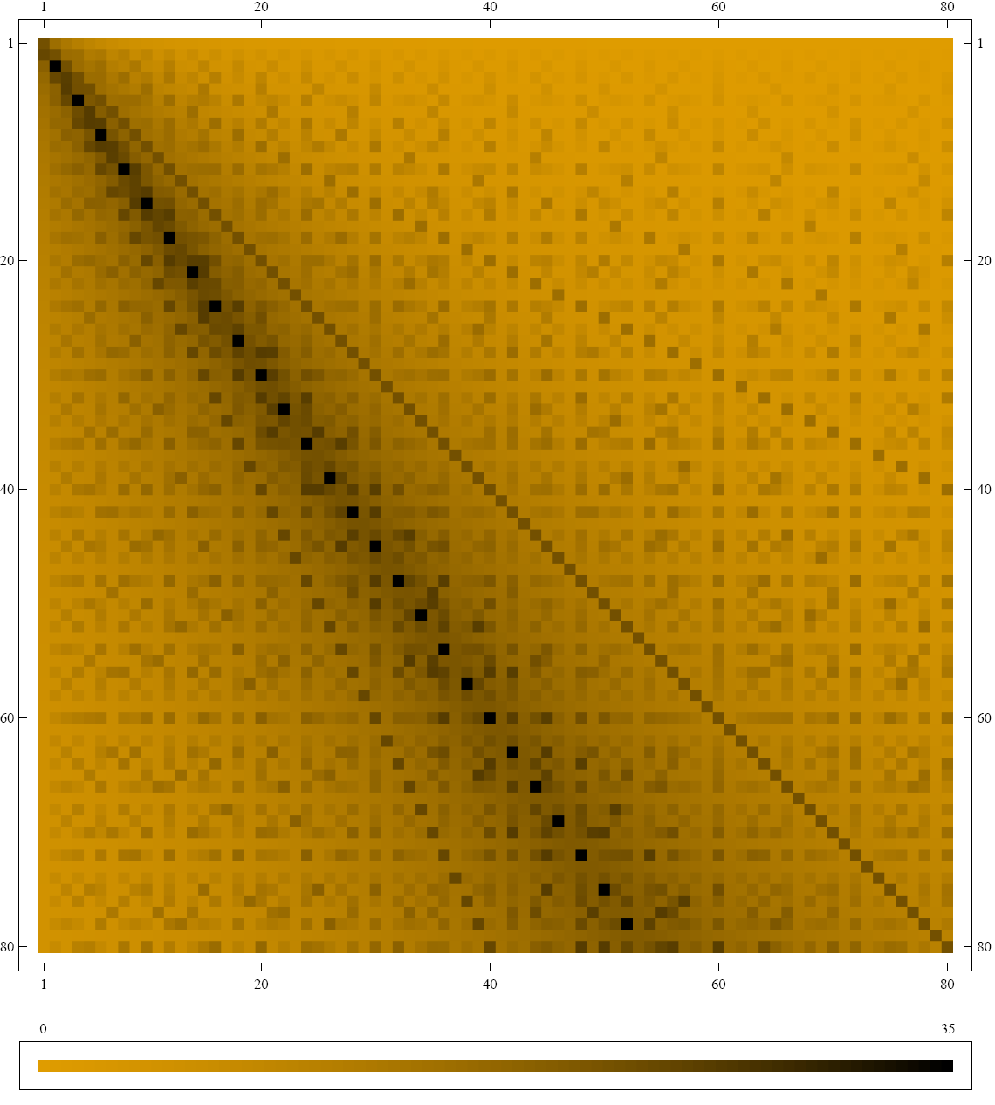}
\caption{Values of $\chi(\alpha,\beta)$ for $1\leq\alpha,\beta\leq80$. The abscissa corresponds to $\beta$. Darker colors represent the largest (better) values of $\chi(\alpha,\beta)$. The best choice is $\alpha=3$, $\beta=2$ for which $\chi(3,2)\approx 35$.}\label{Fig:chi}
\end{figure}
\end{center}

\begin{center}
\begin{figure}[htbp]
\hspace*{0cm}
\includegraphics[width=16cm]{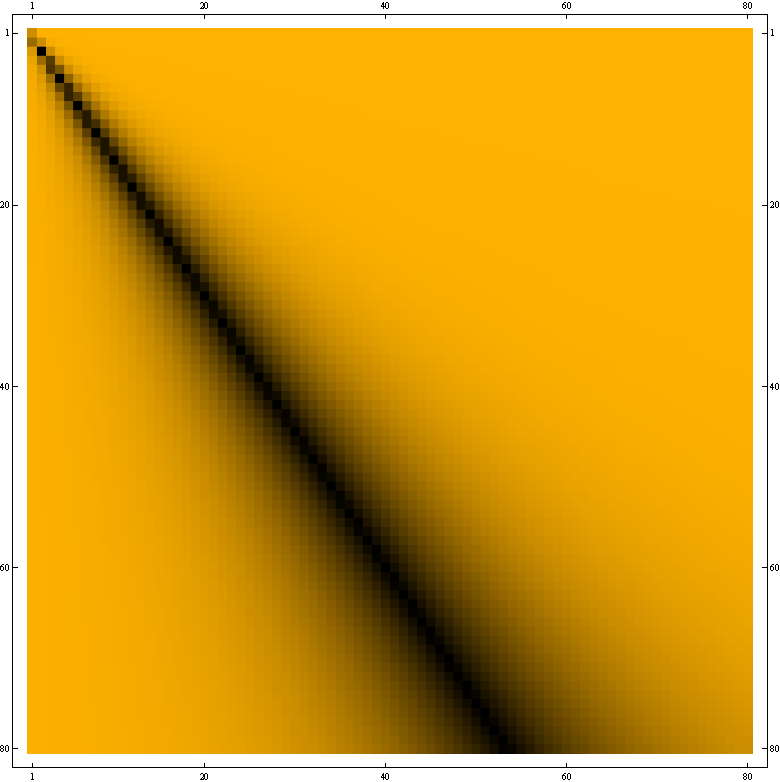}
\caption{Values of $\pi\gamma_{\textrm{stairs}}(\alpha,\beta)$ for $1\leq\alpha,\beta\leq80$. Darker colors represent the largest (better) values of $\pi\gamma_{\textrm{stairs}}(\alpha,\beta)$. The best value corresponds to $\alpha=3$, $\beta=2$ but other choices give values very close to $\pi\gamma_{\textrm{stairs}}(3,2)$. A similar picture can be drawn for $\pi\gamma_{\textrm{gauss}}(\alpha,\beta)$ that essentially coincides with this one for $\alpha>\beta$. When $\alpha<\beta$ the behavior changes because $\pi\gamma_{\textrm{gauss}}(\alpha,\beta)$ becomes larger than $\pi\gamma$ or even complex. This does not change our conclusions about the goodness of $P(3,2)$.}\label{Fig:gamma}
\end{figure}
\end{center}

The values of $P(\alpha,\beta)$ are always positive integers. It is easy to show (see below) that there is always a minimum integer number $p_{\textrm{min}}(\alpha,\beta)$ such that the spacing between consecutive allowed values of  $P(\alpha,\beta)$ is $\gcd(\alpha,\beta)$. Taking this fact into account and using $\nu_0(\alpha,\beta)$, $\mu(\alpha,\beta)$ and $\sigma(\alpha,\beta)$, it is possible to define a set of parameters that can be used to study the goodness of the approximations obtained by using the generalized peak counters introduced here. In particular, we will consider
\begin{eqnarray*}
\chi(\alpha,\beta)&=&\frac{\mu(\alpha,\beta)\sqrt{\gcd(\alpha,\beta)}}{\sigma(\alpha,\beta)}\,,\\
a_{\textrm{min}}(\alpha,\beta)&=&p_{\textrm{min}}(\alpha,\beta)\mu(\alpha,\beta)\,,\\
a_{\textrm{max}}(\alpha,\beta)&=&\gcd(\alpha,\beta)\frac{(1-\chi^2(\alpha,\beta))^2}{4\chi^2(\alpha,\beta)}\mu(\alpha,\beta)
\end{eqnarray*}
and also
\begin{eqnarray*}
\pi \gamma_{\textrm{stairs}}(\alpha,\beta)&=&-\frac{\log\nu_0(\alpha,\beta)}{\mu(\alpha,\beta)}\,,\\
\pi \gamma_{\textrm{gauss}}(\alpha,\beta)&=&\frac{\mu(\alpha,\beta) }{\sigma^2(\alpha,\beta)}\left(1-\sqrt{1+\frac{2\sigma^2(\alpha,\beta)}{\mu ^2(\alpha,\beta)}\log\nu_0(\alpha,\beta)}\right)\,.
\end{eqnarray*}
For our purposes, it suffices to consider those values of $\alpha$ and $\beta$ for which the argument of the square root in $\pi \gamma_{\textrm{gauss}}(\alpha,\beta)$ is positive. By using the following evident facts
\begin{eqnarray*}
& &\nu_0(n\alpha,n\beta)=\nu^{1/n}_0(\alpha,\beta)\,,\quad p_{\textrm{min}}(n\alpha,n\beta)=np_{\textrm{min}}(\alpha,\beta)\,,\\
& & \mu(n\alpha,n\beta)=\frac{1}{n}\mu(\alpha,\beta)\,, \quad \sigma^2(n\alpha,n\beta)=\frac{1}{n}\sigma^2(\alpha,\beta)\,,
\end{eqnarray*}
it is possible to see that these parameters satisfy
\begin{eqnarray*}
& &\chi(n\alpha,n\beta)=\chi(\alpha,\beta)\,,\quad a_{\textrm{min}}(n\alpha,n\beta)=a_{\textrm{min}}(\alpha,\beta)\,,\quad a_{\textrm{max}}(n\alpha,n\beta)=a_{\textrm{max}}(\alpha,\beta)\\
& & \gamma_{\textrm{stairs}}(n\alpha,n\beta)=\gamma_{\textrm{stairs}}(\alpha,\beta)\,,\quad \gamma_{\textrm{gauss}}(n\alpha,n\beta)=\gamma_{\textrm{gauss}}(\alpha,\beta)\,,\quad
\end{eqnarray*}
and, hence, they can be used consistently to assess the appropriateness of the different partitions. Although we will not give here a complete analytic proof, we provide enough numerical evidence to support the choice of $P(3,2)$ as the best peak counter (as was to be expected) in Figs. \ref{Fig:chi} and \ref{Fig:gamma}, where we show the values of $\chi(\alpha,\beta)$ and $\gamma_{\textrm{stairs}}(\alpha,\beta)$ for $1\leq\alpha,\beta\leq80$.
Several comments are in order now. First we want to comment on the role and meaning of $a_{\textrm{min}}(\alpha,\beta)$. It is important to notice that all counters can be used to \textit{exactly} reproduce the behavior of the entropy. This is so because they provide partitions of the set of black hole configurations. However, the behavior of the associated Gaussian approximations --available for all of them-- differ for different choices for $\alpha$, $\beta$ and are not optimal in many instances. For example, before reaching the area value given by $a_{\textrm{min}}(\alpha,\beta)$, it is not true that the Gaussian approximation to the entropy can be understood as the sum of equally spaced Gaussian steps. In fact, the distance between consecutive steps in this regime is dictated by the values of $p$ (non necessarily consecutive) that give non-zero values for $[\nu^p]G_*(\nu,0;\alpha,\beta)$. When $p\geq p_{\textrm{min}}(\alpha,\beta)$ these values are separated by the $\gcd(\alpha,\beta)$ and are ``as consecutive as possible''. When $\gcd(\alpha,\beta)=1$ the value of $p_{\textrm{min}}(\alpha,\beta)$ corresponds to one plus the Frobenius number of the arithmetic sequence $\alpha k+\beta$ with $k\in\mathbb{N}$. This is given by \cite{Roberts} (see also \cite{Alfonsin})
$$
 p_{min}(\alpha,\beta)= 1+\alpha(\alpha+\beta-1)\,.
$$
The value of $a_{\textrm{min}}(\alpha,\beta)$ by itself does not tell us anything about the quality of $P(\alpha,\beta)$ as a counter. In addition to the threshold area, there is another relevant value $a_{\textrm{max}}(\alpha,\beta)$ --that can be roughly defined as the maximum area for which two consecutive steps can be discerned in the gaussian approximation-- that plays an important role. In fact, the interval length $a_{\textrm{max}}(\alpha,\beta)-a_{\textrm{min}}(\alpha,\beta)$ tells us how well the chosen Gaussian approximation captures both the structure at the smallest area scales and the steps in the entropy. Though it is difficult to give a unique definition for $a_{\textrm{max}}(\alpha,\beta)$, it is obvious that the broadening of the Gaussian steps signaled by the growth of the variance $\sigma^2(\alpha,\beta)$ leads to the difficulty of separating two consecutive ones beyond
$$
a_{\textrm{max}}(\alpha,\beta)=\gcd(\alpha,\beta)\frac{(1-\chi^2(\alpha,\beta))^2}{4\chi^2(\alpha,\beta)}\mu(\alpha,\beta)\,.
$$
This condition is derived by requiring that the width of a step is essentially equal to the distance from the previous one. Another quantitative criterion is provided by the exponents in (\ref{decaystar}) and (\ref{decay}) that tell us the decay rate of the staircase structure. The inverse of the numerical coefficient of the area in the exponents of these expressions gives an order of magnitude estimate of $a_{\textrm{max}}(\alpha,\beta)$. In practice the best choice of peak counter is the one giving both a low value of $a_{\textrm{min}}(\alpha,\beta)$ and a large value of $a_{\textrm{max}}(\alpha,\beta)$. The numerical evidence available tells us that the choice $\alpha=3$ and $\beta=2$ is also optimal in this respect.

\subsection{Peak substructures: Using two counters}

\begin{center}
\begin{figure}[htbp]
\hspace*{-.2cm}
\includegraphics[width=16.5cm]{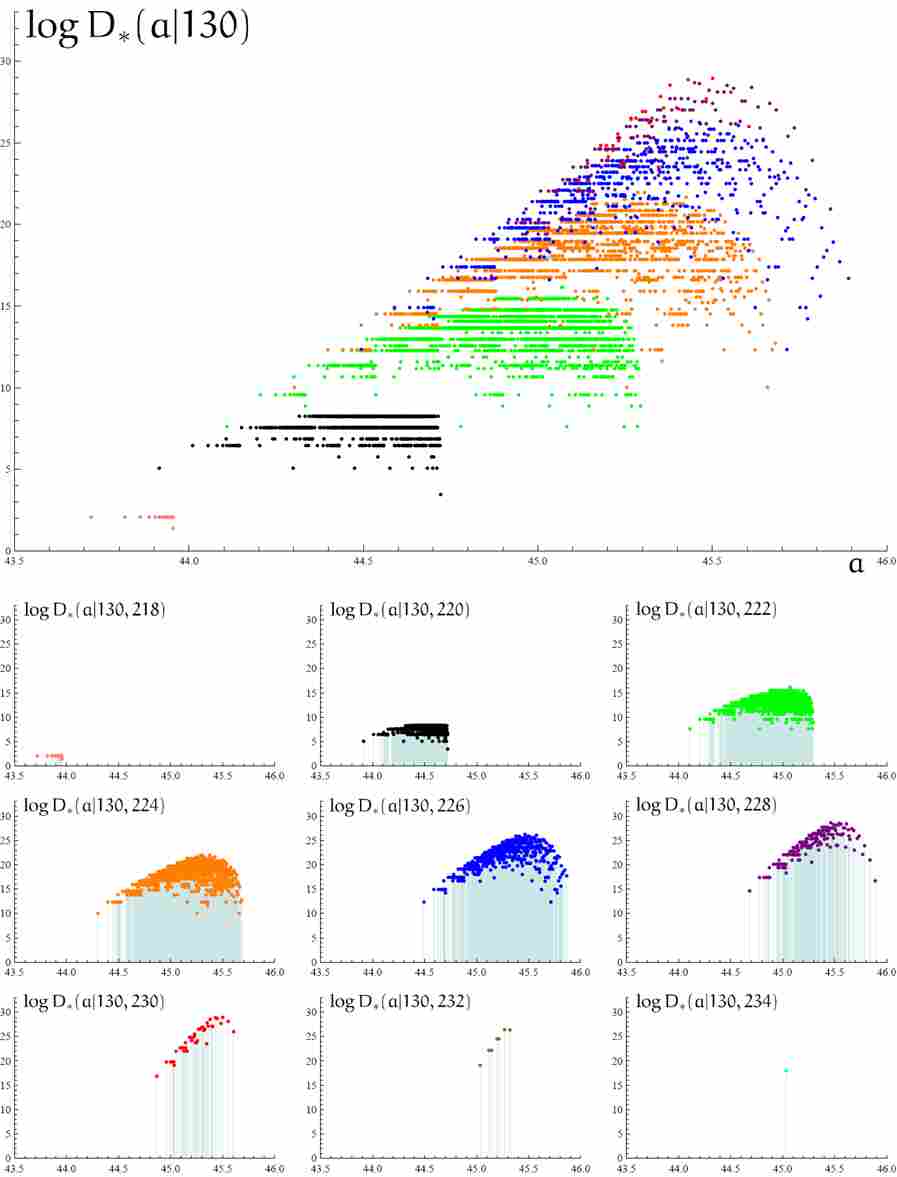}
\caption{Detail of the peak $P_1=3K+2N=130$. This logarithmic plot clearly shows the existence of several ``subpeaks'' labeled by the counter $P_2=5K+4N$. Notice the asymmetric shape (mentioned in Fig. \ref{Fig:peak}).} \label{Fig:subpeaks}
\end{figure}
\end{center}

As we have discussed in the previous section, it is actually possible to partition the configuration space by using different types of linear counters $P(\alpha,\beta)$. The best choice, as far as the description of the entropy structure is concerned, is $P:=3K+2N$. The purpose of this section is to explore the effects of performing a further partition of the configurations corresponding to a single peak $P=p$ by introducing an extra counter. The rationale behind this analysis can be immediately perceived by looking at the structure of a peak (see Fig. \ref{Fig:subpeaks}). As it can be seen, there are coherent substructures within each peak that are responsible for its asymmetric shape in a logarithmic plot. The most obvious and straightforward way to address the description of this substructure is to use an additional peak counter to further partition the space of configurations for a single peak. This can naturally be done by including extra variables in the generating functions. For example, when the projection constraint is ignored,  the moment-generating function
$$
G_*(\nu_1,\nu_2,s):=\frac{1}{1-2\sum_{k=1}^{\infty}\nu_1^{\alpha_1 k+\beta_1}\nu_2^{\alpha_2 k+\beta_2}\exp(-s\sqrt{k(k+2)})}
$$
takes into account the contributions of the counters $P_1=P(\alpha_1,\beta_1)$ and $P_2=P(\alpha_2,\beta_2)$. This moment-generating function satisfies
$$
E_*[\exp(-sA)\,|\, p_1,p_2]=\frac{[\nu_1^{p_1}][\nu_2^{p_2}] G_*(\nu_1,\nu_2,s)}{[\nu_1^{p_1}][\nu_2^{p_2}] G_*(\nu_1,\nu_2,0)}
$$
and can be derived from the obvious generalization of the master generating function (\ref{master*})
\begin{eqnarray}
  G_*(\nu_1,\nu_2;x_1,x_2,\dots)&:=&\left(1-2\sum_{i=1}^\infty \sum_{ n=1}^\infty  \nu_1^{\alpha_1 k_ n ^i +\beta_1}\nu_2^{\alpha_2 k_ n ^i +\beta_2} x_i^{y_ n^i}\right)^{-1}\,.\label{master_2*}
\end{eqnarray}
The black hole degeneracy spectrum for the configurations satisfying $P_1=p_1$, $P_2=p_2$ and area $a=q_1\sqrt{p_1}+q_2\sqrt{p_2}+\cdots$ is given by
$$
D_*(a\,|\,p_1,p_2)=[x_1^{q_1}x_2^{q_2}\cdots ][\nu_1^{p_1}][\nu_2^{p_2}] G_*(\nu_1,\nu_2;x_1,x_2,\dots)\,.
$$
In Fig. \ref{Fig:subpeaks} we use $P_1:=P(3,2)=3K+2N$ and $P_2:=P(5,4)=5K+4N$. This choice is favored by the arguments presented in \cite{Agullo:2008eg} and can also be understood by looking at the role of the Pell equation in the characterization of the area spectrum. A statistical description of the subpeaks can be made by following the steps discussed in previous sections. In particular a Gaussian approximation can be obtained that, presumably, improves the one given by a single peak counter $P(3,2)$. However, as we do not expect the problems encountered above to be fixed by following this approach (in particular the underestimate of the value of $\gamma$), we will not pursue it further. In any case, it is obvious that, by introducing extra counters and studying the statistical properties of the resulting peaks, one can get improved approximations for the entropy.

\section{Conclusions and comments}\label{conclusions}

The main idea of the paper is to use the master generating functions, that encode the black hole degeneracy spectrum in an exact way, to derive statistical properties that can be used to describe and understand the detailed features observed in the black hole entropy. We have succeeded in reproducing, from a purely analytical point of view, the staircase of the entropy and its behavior as a function of the horizon area. In particular, we have shown how and why the steps disappear. The key element of our approach has been to use statistical properties of some subsets of black hole configurations (the ``peaks'' defined by $P=3K+2N$) to construct analytic approximations for the black hole entropy. It is very important to highlight the fact that we are not merely fitting the data but, rather, computing, \textit{a priori}, the relevant statistical parameters of the peaks by employing the BH generating functions. In particular, we have given a procedure that can be used to get all the moments of the area distribution.

It is also necessary to emphasize that the partitions of the space of BH configurations that we have exploited are \textit{exact}, so there is no approximation introduced by the choice of the different ``peak counters''. The smoothing of the entropy consisting on adding Gaussian steps is a natural one that works remarkably well for small areas. Furthermore, there are important theorems in Combinatorics that guarantee the convergence of the individual steps in the entropy (selected by the peak counter) to Gaussian distributions (after normalization). However, the sum of the Gaussian steps does not converge to the entropy because, as we have shown, the linear growth predicted the Gaussian approximation is slightly smaller than the actual one. It must be pointed out here that the numerical estimates for $\gamma$, even in the crudest approximations, are remarkably good. In any case, the area range where the Gaussian approximation is reliable (that can be essentially obtained by comparing the actual growth given by the true $\gamma$ and $\gamma_{\textrm{gauss}}$) is large enough to trust the Gaussian approximation regarding the disappearance of the staircase structure (see Fig. \ref{Fig:esc_gauss_3}). Finally, we have discussed in Appendix \ref{App:C} an alternate way of assessing the validity of our approach by looking at the pole structure of the Laplace transforms of the entropy and its Gaussian approximation. The comparison of both analytic structures tells us that they differ in their behavior far from the real axis. The numerical evidence available gives tantalizing evidence (encoded in an approximate periodicity that can be readily seen in Fig. \ref{Fig:compare}), that prevents us from excluding a revival of the observed staircase structure for large area values. In any case, we deem this possibility quite unlikely.

We have given numerical evidence to support the election of $P=3K+2N$ as the best peak counter within the class $P(\alpha,\beta)=\alpha K+\beta N$, $\alpha,\beta\in \mathbb{N}$. Its usefulness is justified by the fact that in the low area regime the variance of the distributions associated with the peaks is much smaller than the separation of the mean areas corresponding to consecutive peaks. This explains why a staircase structure must be seen in this regime. The actual comparison of the exact entropy values and the prediction given by our model is very compelling. In any case, it is conceivable that other functions, more general than the counters $P(\alpha,\beta)$ that we have discussed, can be defined in order to better understand and approximate the large area behavior of the entropy and the value of $\gamma$.

Although most of the computations carried out in the paper have made use of the DL prescription to obtain the black hole entropy \cite{Domagala:2004jt}, our methods can be extended in a completely straightforward way to other countings such as the $SU(2)$ proposal of \cite{Engle:2009vc,Engle:2010kt}. The numerical details differ from the ones that we have presented above but the qualitative conclusions  --which are independent of the concrete form of the projection constraint or equivalent conditions-- remain unchanged. In particular, when the condition that plays the role of the projection constraint in this case is ignored, the moment-generating function is \cite{Agullo:2010zz}
$$
G^{\textrm{ENP}}_*(\nu,s)=\frac{1}{1-\sum_{k=1}^\infty (k+1)\nu^{3k+2}\exp(-s\sqrt{k(k+2)})}\,.
$$
By following the methods described in the paper step by step, one finds that $\nu_0=(0.73996900\cdots)$ is now the smallest real root of the polynomial $\nu ^8+\nu ^6-2 \nu ^5-2 \nu ^3+1$. The values for $\mu$ and $\sigma^2$ are obtained from the function
$$
Q^{\textrm{ENP}}_*(\nu,s)=1-\sum_{k=1}^\infty (k+1)\nu^{3k+2}\exp(-s\sqrt{k(k+2)})\,.
$$
They are $\mu=(0.34980945\cdots)$ and $\sigma^2=(0.00010926\cdots)$. The staircase and Gaussian approximations to the entropy lead to the following values for the Immrizi parameter: $\pi\gamma_{\textrm{stairs}}=(0.86088860\cdots)$ and $\pi\gamma_{\textrm{gauss}}=(0.86100438\cdots)$. These have to be compared with $\pi\gamma_{\textrm{ENP}}=(0.86100642\cdots)$ obtained in \cite{Agullo:2009eq}. The analytic structure of the Gaussian approximation is similar to the one shown in Fig. \ref{Fig:sing} and, then, the behavior of the entropy is essentially the same as in the DL case. It is obtained from this one by changing the values of the parameters $\nu_0$, $\mu$ and $\sigma^2$. The projection constraint can be introduced in the same way as before and, as expected, only configurations with even $P$ contribute in this case. The relevant singularities in the Laplace transform are located at $s_{*\pm}=(0.81057903\cdots)\mp (17.96628553\cdots) i$ and $s_{\pm}=(0.84839767\cdots)\mp (8.98324891\cdots) i$. This leads to a doubling of the size of the steps and the persistence of the staircase structure for larger values of the horizon area in the case when the projection constraint is taken into account.

A last comment that we want to add is related to the values of the areas for which the steps in the entropy and the peaks in the BH degeneracy spectrum cease to be seen. As the entropy is obtained by integrating the degeneracy spectrum and taking the logarithm of the resulting sum, it is to be expected that the effective disappearance of the steps takes place for smaller values of the areas
than the disappearance of the corresponding peaks in the degeneracy spectrum. This must be taken into account in order to correctly interpret the meaning of the substructures found in the behavior of the entropy as a function of the area.

\acknowledgments

We wish to thank Ivan Agullo, Jacobo Diaz-Polo and Enrique F. Borja for their very interesting comments and encouragement. We also want to thank Jes\'us Salas for patiently listening to our comments on the paper and many helpful discussions. Finally we also thank Pablo San Jos\'e for his invaluable assistance with optimizing our \textit{Mathematica}$^{TM}$ codes. This work has been supported by the Spanish MICINN research grant FIS2009-11893 and  the  Consolider-Ingenio 2010 Program CPAN (CSD2007-00042). Some of the computations and plots have been done with the help of \textit{Mathematica}$^{TM}$.

\appendix

\section{Computation of the moments when the projection constraint is not considered}\label{App:A}
In this Appendix we will derive the mean and variance of the area distribution in the non-projection constraint setting by using (\ref{momentgf}).
They are given by
\begin{eqnarray*}
\mu_{*p}&:=&E_*[A\,|\,p]=\frac{2}{\alpha_{*p}}[\nu^p]\left(\left(\frac{\nu^3-1}{2\nu^5+\nu^3-1}\right)^2\sum_{k=1}^\infty \sqrt{k(k+2)}\nu^{3k+2}\right)
\end{eqnarray*}
and $\displaystyle \sigma_{*p}^2=E_*[A^2\,|\,p]-\mu_{*p}^2$,
where
$$
E_*[A^2\,|\,p]=\frac{1}{\alpha_{*p}}[\nu^p]\left(\frac{2 \nu^5 (\nu^3-3)}{(\nu^3-1)(2\nu^5+\nu^3-1)^2}+
\frac{8(\nu^3-1)^3\Big(\sum_{k=1}^\infty \sqrt{k(k+2)}\nu^{3k+2}\Big)^2}{(2\nu^5+\nu^3-1)^3}\right)\,.
$$
Here, as in the main body of the paper, we use the notation
$$
\alpha_{*p}:=[\nu^p]\left(\frac{\nu^3-1}{2\nu^5+\nu^3-1}\right)\,.
$$
The computation of the coefficients in the Taylor expansions about $\nu=0$ that appear in the previous formulas can be easily carried out, for instance, by using the Cauchy integral theorem
$$
[\nu^p]f(\nu)=\frac{1}{2\pi i}\oint_C \frac{f(\nu)}{\nu^{p+1}}\,\mathrm{d}\nu,
$$
where $C$ is an index-one curve surrounding the origin $\nu=0$ (and leaving the remaining singularities of the integrand outside). The pole at the origin has order $p+1$ so, in practice, it is better to compute the integral by moving the contour radially outwards and picking up the contributions of the remaining singularities of $f(\nu)/\nu^{p+1}$. This is useful because they are, in many cases, poles of a fixed, $p$-independent order with a $p$-dependent residue. The value of $\alpha_{*p}$ can be easily obtained by using this procedure:
\begin{eqnarray*}
\alpha_{*p}&=&\frac{1}{2\pi i}\oint_C \frac{1}{\nu^{p+1}}
       \frac{\nu^3-1}{2\nu^5+\nu^3-1}\,\mathrm{d}\nu\\
       &=&-\sum_{i=-2}^2 \mathrm{Res}\left(\frac{\nu^3-1}{\nu^{p+1}(2\nu^5+\nu^3-1)};\nu_i\right)=\sum_{i=-2}^2\frac{2\nu_i^2}{(10\nu_i^2+3)}\frac{1}{\nu_i^p}
\end{eqnarray*}
where the $\nu_i$ are the five different roots of the polynomial $2\nu^5+\nu^3-1$. We have used the convention that $\nu_i$ and $\nu_{-i}$ are complex conjugate of each other. The single real root, $\nu_0=(0.77039825\cdots)$, is the one with the smallest module (a closed expression for $\nu_0$ in terms of hypergeometric functions is given in \cite{Agullo:2010zz}).  The previous expression shows that the asymptotic behavior for large values of $p$ of $\alpha_{*p}$ is given by
$$
\alpha_{*p}\sim \frac{2\nu_0^2}{(10\nu_0^2+3)}\frac{1}{\nu_0^p}\,,\quad p\rightarrow \infty.
$$
There are other alternative expressions for the coefficients $\alpha_{*p}$. For example they satisfy the simple recursion relation
$$
\alpha_{*p+5}=\alpha_{*p+2}+2\alpha_{*p},\quad p\geq0,\quad \alpha_{*0}=1,\alpha_{*1}=\alpha_{*2}=\alpha_{*3}=\alpha_{*4}=0\,.
$$
They can also be written in terms of binomial coefficients as
$$
\alpha_{*p}=\sum_{n=\lfloor\frac{p+1}{3} \rfloor}^{\lfloor\frac{2p}{5} \rfloor} 2^{3n-p}\binom{p-2n-1}{3n-p-1}
\,,\quad p\geq 5\,.
$$
The numerators appearing in the expressions for the mean and the variance can also be obtained in closed form by using the Cauchy integral formula. However the expressions thus obtained are not very illuminating so, here we will only give their leading asymptotic behavior for large values of $p$. In both cases this is given by a term linear in $p$:
\begin{eqnarray*}
\mu_{*p}&\sim& \mu \cdot p + \mu_0= (0.34959022\cdots)\cdot p+(0.00019724\cdots)\,\\
\sigma_{*p}^2&\sim& \sigma^2 \cdot p = (0.00009817\cdots) \cdot p\,,
\end{eqnarray*}
where $\mu$,   $\mu_0$, and $\sigma^2$ are given in terms of $\nu_0$:
\begin{eqnarray*}
\mu&=&\frac{4\nu^2_0}{10\nu_0^2+3} \left(\sum_{k=1}^\infty \sqrt{k(k+2)}\nu_0^{3k+2}\right)\\
\mu_0&=&\left(\frac{8\nu_0^2(20\nu^2_0+3)}{(10\nu^2_0+3)^2}+\frac{2(5\nu^3_0+1)}{\nu^3_0(10\nu^2_0+3)}\right)\sum_{k=1}^\infty \sqrt{k(k+2)}\nu_0^{3k+2}\\
&-&
 \frac{2\nu_0(1-\nu_0^3)}{\nu^3_0(10\nu^2_0+3)}\sum_{k=1}^\infty (3k+2)\sqrt{k(k+2)}\nu_0^{3k+1}\\
\sigma^2&=& \beta-2\mu\cdot \mu_0\,,
\end{eqnarray*}
where
\begin{eqnarray*}
\beta&=&
\frac{3 - \nu_0^3}{\nu_0^3 (1 - \nu_0^3) (10 \nu_0^2 + 3)}\\
&+& \left(\frac{12 (20 \nu_0^2 + 3)(1 - \nu_0^3)^3 }{\nu_0^{11} (10 \nu_0^2 + 3)^3 }
           + \frac{6 (1 - \nu_0^3)^2 (5 \nu_0^3 + 1)}{\nu_0^{11} (10 \nu_0^2 + 3)^2 }\right) \left(\sum_{k=1}^\infty \sqrt{k(k+2)}\nu_0^{3k+2}\right)^2
    \\
    &-&\frac{ 8 \nu_0 (1 - \nu_0^3)^3}{\nu_0^{11} (10 \nu^2_0 + 3)^2}\left(\sum_{k=1}^\infty \sqrt{k(k+2)}\nu_0^{3k+2}\right)\left(\sum_{k=1}^\infty (3k+2)\sqrt{k(k+2)}\nu_0^{3k+1}\right)
    \vspace*{5mm}\\
    &=& (0.00023608\cdots)\,.
\end{eqnarray*}
The results obtained by this method are completely equivalent to those presented in Section \ref{gaussian}. Higher moments can be computed by using the same procedure described here.

\section{Computation of the moments when the projection constraint is considered}\label{App:B}

As we have discussed in subsection \ref{DL}, the incorporation of the projection constraint requires us to introduce of an extra variable $z$ in the moment-generating function $G(\nu,s;z)$. To compute the moments in the area distribution we have find out some $s$-derivative, evaluated at $s=0$, of $G(\nu,s;z)$. This derivative gives rise to a new function $f(\nu,z)$ whose coefficient $f_p=[z^0][\nu^p]f(z,\nu)$ allows us to determine the moments in the area distribution. This process defines a function $F(\nu)=\sum_p f_p \nu^p$ of the single variable $\nu$ that is desirable to have in closed form. We describe in this subsection how this can be done. What we need to find is
$$
F(\nu)=\sum_{p=0}^\infty f_p \nu^p=\sum_{p=0}^\infty \nu^p[z^0][\tilde{\nu}^p]f(\tilde{\nu},z)=\sum_{q=0}^\infty \nu^p \frac{1}{2\pi i}\oint_{C_1}\frac{\mathrm{d}z}{z}[\tilde{\nu}^p]f(\tilde{\nu},z)
$$
whenever $f(\tilde{\nu},z)$ is an analytic function of $z$ for all the values of $\tilde{\nu}$ in a neighborhood of $\tilde{\nu}=0$. If $[\tilde{\nu}^q]f(z,\tilde{\nu})$ is a Laurent polynomial at $z=0$ we can choose any contour surrounding $z=0$ for $C_1$ (as long as it is piecewise smooth and of index one). Now if $f(\tilde{\nu},z)$ is analytic at $\tilde{\nu}=0$ for all the values of $z\in C_1$ the previous expression can be rewritten as
$$
F(\nu)=\sum_{p=0}^\infty \nu^p \frac{1}{2\pi i}\oint_{C_1}\frac{\mathrm{d}z}{z} \frac{1}{2\pi i}\oint_{C_2} \frac{\mathrm{d}\tilde{\nu}}{\tilde{\nu}^{p+1}}  f(\tilde{\nu},z)\,,
$$
where the contour $C_2$ (again piecewise smooth and of index one) is now chosen in such a way that, for each $z\in C_1$  the only singularity surrounded by it is $\tilde{\nu}=0$. We can now exchange the order in the integrals to get
\begin{equation}
F(\nu)=\sum_{p=0}^\infty \nu^p \frac{1}{2\pi i}\oint_{C_2} \frac{\mathrm{d}\tilde{\nu}}{\tilde{\nu}^{p+1}} \frac{1}{2\pi i}\oint_{C_1}\frac{\mathrm{d}z}{z}  f(\tilde{\nu},z)=\frac{1}{2\pi i}\oint_{C_1}\frac{\mathrm{d}z}{z}  f(\nu,z)\,,
\label{integral}
\end{equation}
whenever the last integral is an analytic function of $\nu$ in an open neighborhood of $\nu=0$. If the integration contours are chosen according to the prescription described here, there are (generically) singularities of $z\mapsto f(\nu,z)/z$ inside $C_1$ (in most of the cases $z=0$ and, eventually,  others coming from $f(\nu,z)$; which will be functions of $\nu$). The residues at these singularities give us a closed-form expression for
$$
F(\nu)=\frac{1}{2\pi i}\oint_{C_1}\frac{\mathrm{d}z}{z}  f(\nu,z)\,.
$$
In many practical situations it is convenient to choose  a unit, positively oriented, circumference for $C_1$. This is so because, for this choice, it is possible to simplify expressions of the type
$$
F(\nu)=\frac{1}{2\pi i}\oint_{C_1} \frac{dz}{z} (z^k+z^{-k}) g(\nu,z+z^{-1})=\frac{1}{\pi i}\oint_{C_1} \frac{dz}{z} z^k g(\nu,z+z^{-1})\,,\quad k\in \mathbb{N},
$$
as can be easily seen by performing the change of variable $z\mapsto z^{-1}$ in the second term of the first integral.
In the particular case of interest, by setting $s=0$ in the master generating function $G(\nu,s;z)$, we find
\begin{eqnarray*}
f(\nu,z)&=&G(\nu,0;z)=\frac{1}{1-\sum_{k=1}^\infty (z^k+z^{-k})\nu^{3k+2}}
=\frac{z^2 \nu ^3-z \nu ^6-z+\nu ^3}{z^2 \nu ^5+z^2 \nu ^3-2 z \nu ^8-z \nu ^6-z+\nu ^5+\nu ^3}\,.
\end{eqnarray*}
In order to apply the previous procedure we first take a unit, positively oriented, circumference for $C_1$. With this choice, and once a suitable $C_2$ contour is picked, the two relevant poles in the integrand (\ref{integral}) are $z=0$ and
$$
z=\frac{1+2 \nu ^8+\nu ^6-\sqrt{\left(\nu ^2-1\right) \left(\nu ^4+\nu ^2+1\right) \left(2 \nu ^5+\nu ^3-1\right) \left(2 \nu ^5+\nu^3+1\right)}}{2\nu^3 \left(\nu ^2+1\right)}\,,
$$
so that
$$
F(\nu)=\frac{1}{1+\nu^2}\left(1+\nu^2\sqrt{\left(\frac{\nu^3-1}{2\nu^5+\nu^3-1}\right)
\left(\frac{\nu^3+1}{2\nu^5+\nu^3+1}\right)}\,\right)\,.
$$
Some facts are evident at this point. In particular, the coefficient $
\alpha_p:=[\nu^p]F(\nu)$
is zero for odd values of $p$. For large even values of $p$ the asymptotic behavior of $\alpha_p$ is controlled by the singularities $\nu=\pm \nu_0$ and is
$$
\alpha_{2q}\sim\frac{1}{1+\nu_0^2}\sqrt{\frac{2\nu_0(1-\nu_0^6)}{\pi (10\nu_0^2+3)}}\frac{1}{\sqrt{2q}\nu_0^{2q}}\,.
$$

The mean $\mu_p$ is computed by following the same steps as in the non-projection constraint case. In particular $\alpha_p\cdot \mu_p$ is given by the $[\nu^p]$ coefficient of
\begin{eqnarray*}
& & \frac{2\nu^6(\nu^6-1)^2}{(1+\nu^2)^2R(\nu)}\sum_{k=1}^{\infty}k\sqrt{k(k+2)}\left(\frac{2\nu^6 (1+\nu^2)}{Q_1(\nu)+\sqrt{R(\nu)}}\right)^k \\
& & +\frac{2\nu^4(\nu^6-1)^2Q_2(\nu)}{(1+\nu^2)^2R(\nu)\sqrt{R(\nu)}}\sum_{k=1}^{\infty}\sqrt{k(k+2)}\left(\frac{2\nu^6 (1+\nu^2)}{Q_1(\nu)+\sqrt{R(\nu)}}\right)^k\,,
\end{eqnarray*}
where
\begin{eqnarray*}
R(\nu)&:=&(\nu^2-1)(\nu^4+\nu^2+1)(2\nu^5+\nu^3+1)(2\nu^5+\nu^3-1)\\
Q_1(\nu)&:=&2\nu^8+\nu^6+1\\
Q_2(\nu)&:=&2+\nu^2-2\nu^6-7\nu^8-6\nu^{10}\,.
\end{eqnarray*}
For large even values of $p$, the asymptotic behavior of $\mu_p$ is also controlled by the singularities $\nu=\pm \nu_0$. In particular, by using the identities
$$
\left(\frac{2\nu_0^6(1+\nu_0^2)}{2\nu_0^8+\nu_0^6+1+\sqrt{R(\pm\nu_0)}}\right)^k
=\left(\frac{2\nu_0^6(1+\nu_0^2)}{2\nu_0^8+\nu_0^6+1}\right)^k
=\nu^{3k}_0\,,
$$
it is straightforward  to show that
$$
\mu_{2q}\sim 2\mu q\,,\quad q\rightarrow\infty\,,
$$
where, as in the non-projection constraint case,
$$
\mu=\frac{4\nu^2_0}{10\nu_0^2+3} \left(\sum_{k=1}^\infty \sqrt{k(k+2)}\nu_0^{3k+2}\right)\,.
$$
Finally, $(\sigma_p^2+\mu_p^2)\cdot \alpha_p$ is given by the $[\nu^p]$ coefficient of
\begin{eqnarray*}
&&\frac{4\nu^{10}(\nu^6-3)}{(1+\nu^2)^3(\nu^6-1)^3}+\frac{4\nu^{12}(\nu^6-3)Q_5(\nu)}{(1+\nu^2)^3(\nu^6-1)R(\nu)^2\sqrt{R(\nu)}}+
\frac{4(1+\nu^2)(\nu^6-1)\Big(Q_6(\nu)\sqrt{R(\nu)}+Q_4(\nu)\Big)}{\nu^2R(\nu)\Big(Q_3(\nu)\sqrt{R(\nu)}-Q_1(\nu)R(\nu)\Big)}\\
&&+\frac{(\nu^6-1)^3\nu^6}{(1+\nu^2)^3R^2(\nu)\sqrt{R(\nu)}}\sum_{k_1=1}^\infty\sum_{k_2=1}^{k_1-1}2^{2+k_2-k_1}\sqrt{k_1(k_1+2)k_2(k_2+2)}
\left(\frac{Q_1(\nu)-\sqrt{R(\nu)}}{1+\nu^2}\right)^{k_1-k_2}\nu^{6k_2}\times\\
&&\hspace{7cm}\left(Q_7(\nu)+3(k_2-k_1)\nu^2 Q_2(\nu)\sqrt{R(\nu)}-(k_2-k_1)^2\nu^4 R(\nu)\right)\\
&&+\frac{(\nu^6-1)^3\nu^6}{(1+\nu^2)^3R^2(\nu)\sqrt{R(\nu)}}\sum_{k_1=1}^\infty\sum_{k_2=1}^\infty2^{1-k_1-k_2}\sqrt{k_1(k_1+2)k_2(k_2+2)}
\left(\frac{Q_1(\nu)-\sqrt{R(\nu)}}{1+\nu^2}\right)^{k_1+k_2}\times\\
&&\hspace{7cm}\left(Q_7(\nu)-3(k_1+k_2)\nu^2 Q_2(\nu)\sqrt{R(\nu)}-(k_1+k_2)^2\nu^4 R(\nu)\right)
\end{eqnarray*}
where
\begin{eqnarray*}
Q_3(\nu)&:=&1\!-\!2\nu^{10}\!+\!\nu^{12}\!+\!4\nu^{14}\!+\!4\nu^{16}\\
Q_4(\nu)&:=&1\!-\!\nu^2\!+\!\nu^4\!-\!3\nu^6\!-\!3\nu^8\!-\!5\nu^{10}\!+\!3\nu^{12}\!+\!3\nu^{14}\!+\!11\nu^{16}\!+\!7\nu^{18}\!+\!7\nu^{20}\!-\!\nu^{22}\!-\!4\nu^{26}\\
Q_5(\nu)&:=&3\!+\!3\nu^2\!+\!\nu^4\!-\!9\nu^6\!-\!27\nu^8\!-\!27\nu^{10}\!+\!5\nu^{12}\!+\!47\nu^{14}\!+\!81\nu^{16}\!+\!65\nu^{18}\!+\!5\nu^{20}\!-\!55\nu^{22}\!-\!64\nu^{24}\!-\!28\nu^{26}\\
Q_6(\nu)&:=&\!-\!1\!+\!\nu^2\!-\!\nu^4\!+\!2\nu^6\!+\!2\nu^8\!+\!4\nu^{10}\!-\!\nu^{12}\!+\!\nu^{14}\!-\!\nu^{16}\!+\!2\nu^{18}\\
Q_7(\nu)&:=&\!-\!2(3\!+\!3\nu^2\!+\!\nu^4\!-\!6\nu^6\!-\!24\nu^8\!-\!26\nu^{10}\!-\!\nu^{12}\!+\!23\nu^{14}\!+\!55\nu^{16}\!+\!64\nu^{18}\!+\!28\nu^{20})\,.
\end{eqnarray*}
The asymptotic behavior of $\sigma_{2q}$ can be obtained in a straightforward (albeit tedious) way from the preceding expressions.

\section{Analytical properties of the Laplace transform of the entropy in the Gaussian approximation}\label{App:C}

In order to understand the main features of the Gaussian approximation for the entropy, in particular its asymptotic behavior as a function of the area, it is very convenient to rely on Laplace transform methods. This is so because the asymptotics of a function can be understood in many cases by looking at the analytic structure its Laplace transform, specifically the locations of its singularities and their type. In our case, and despite the fact that the behavior of the Gaussian approximation can be roughly understood from the features of the individual steps, the detailed behavior of the sum is much harder to get. In particular the realization of the fact that the value of the Immirzi parameter is not exactly recovered in this approximation really demands a detailed analysis for which the use of Laplace transform methods is especially appropriate. The arguments given below refer to $P(3,2)$ but can be trivially extended for other peak counters $P(\alpha,\beta)$, defined in Section \ref{sect:substructure}, as long as
$$
1+\frac{2\sigma^2(\alpha,\beta)}{\mu ^2(\alpha,\beta)}\log\nu_0(\alpha,\beta)>0\,.
$$
\begin{center}
\begin{figure}[htbp]
\hspace*{-.8cm}
\includegraphics[width=17cm]{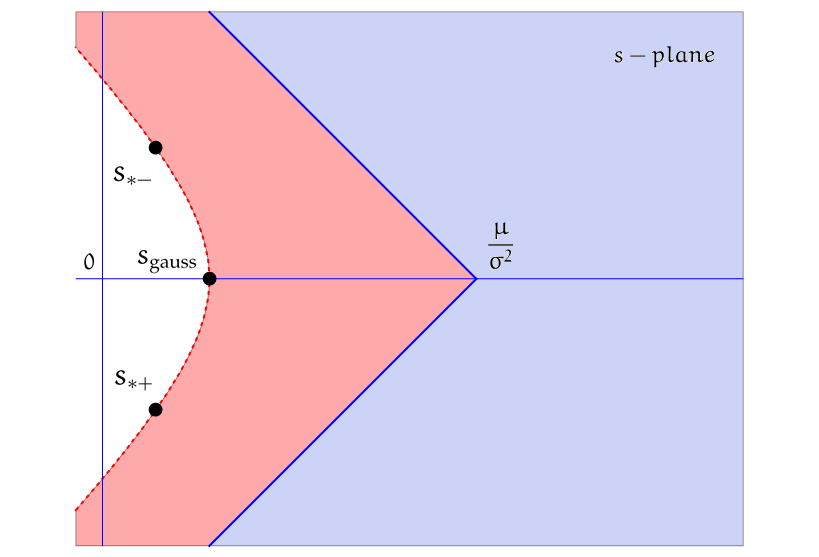}
\caption{We represent here the regions relevant to the discussion of the convergence of the functional series (\ref{series}). We also plot the position of the relevant singularities of the Laplace transform (there is an infinite number of other isolated singularities also located on the hyperbola which are not relevant for the analysis presented in the paper).} \label{Fig:sing}
\end{figure}
\end{center}
\begin{center}
\begin{figure}[htbp]
\hspace*{-.2cm}
\includegraphics[width=16cm]{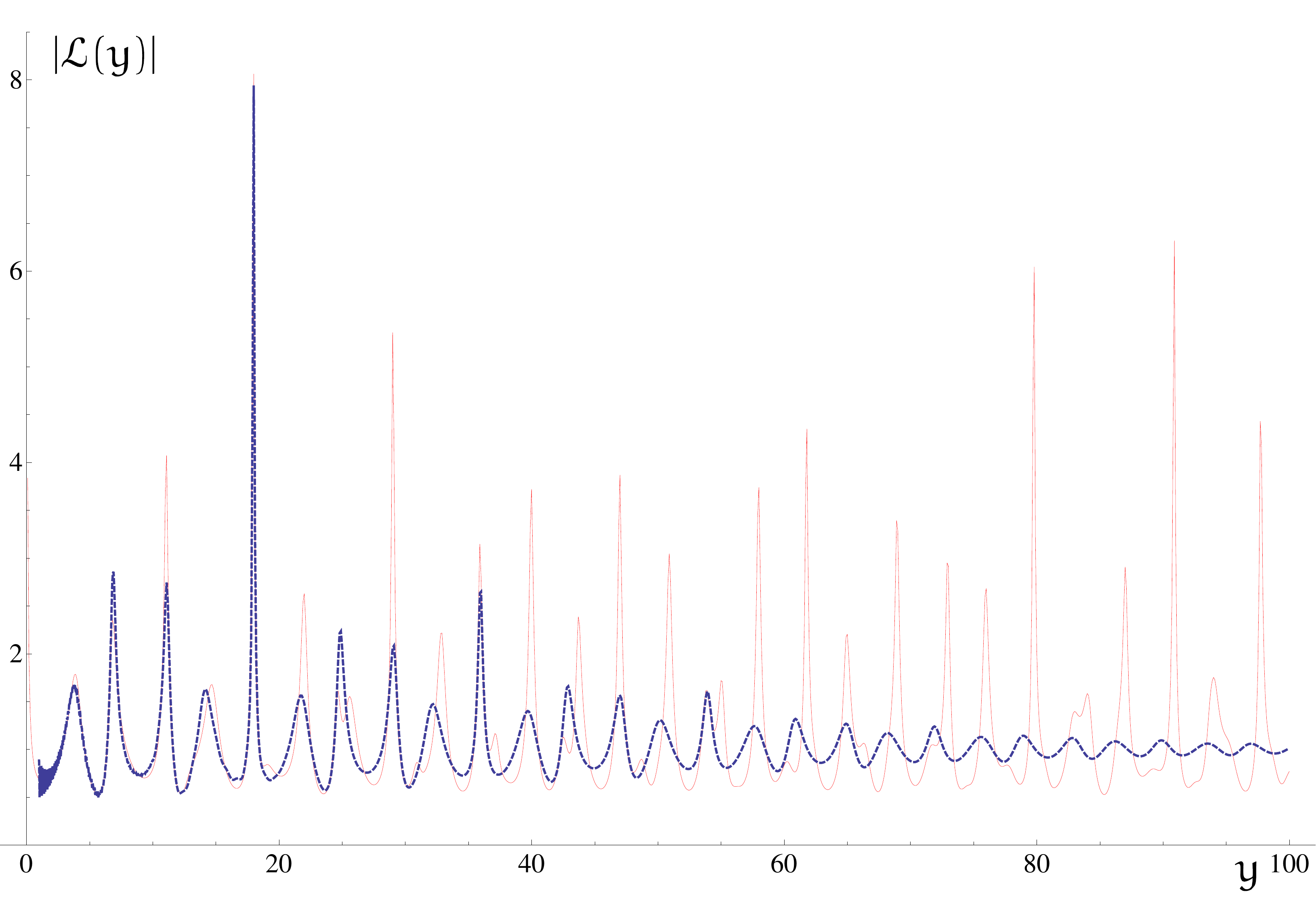}
\caption{Comparison between the absolutes values of the Laplace transforms $\mathcal{L}(\exp S_*,\pi\gamma+i y)$ (continuous line) and $\mathcal{L}(\exp S^{\textrm{gauss}}_*,\pi\gamma+i y)$ (dashed line).}
\label{Fig:compare}
\end{figure}
\end{center}

Let us consider first the non-projection constraint case
\begin{eqnarray}
\mathcal{L}(\exp S^{\textrm{gauss}}_*,s)&=&\frac{1}{s}
+\frac{\nu_0^2}{(10\nu_0^2+3)s}\sum_{p=5}^{\infty} \frac{1}{\nu_0^p}\left(1-\mathrm{erf}\left(\frac{\mu}{\sqrt{2}\sigma }\sqrt{p}\right)\right)\label{laplaceappendix}\\
&+&\frac{\nu_0^2}{(10\nu_0^2+3)s}\sum_{p=5}^{\infty} \frac{1}{\nu_0^p}
\left(1+\mathrm{erf}\left(\frac{\sigma^2s-\mu}{\sqrt{2}\sigma }\sqrt{p}\right)\right)
\exp\left(\Big(\frac{\sigma^2}{2}s^2-\mu s\Big)p\right)\nonumber\,.
\end{eqnarray}
The well known asymptotic behavior of the error function
$$
\mathrm{erf}(x)\sim 1-\frac{e^{-x^2}}{\sqrt{\pi}x}\,,\quad x\gg 1\,,
$$
guarantees the convergence of the first series in (\ref{laplaceappendix}), hence, the first to terms  have a very simple singularity structure: just a simple pole at $s=0$. Let us concentrate then in the last series in (\ref{laplaceappendix})
\begin{equation}
\sum_{p=5}^{\infty} \frac{1}{\nu_0^p}
\left(1+\mathrm{erf}\left(\frac{\sigma^2s-\mu}{\sqrt{2}\sigma }\sqrt{p}\right)\right)
\exp\left(\Big(\frac{\sigma^2}{2}s^2-\mu s\Big)p\right)\,.
\label{series}
\end{equation}
By using now the following asymptotic formula for the error function for complex values of its argument
$$
e^{z^2}(1-\mathrm{erf}(z))\sim \frac{1}{\sqrt{\pi}z}\left(1+\sum_{m=1}^\infty (-1)^m\frac{(2m-1)!!}{(2z^2)^m}\right)\,,\quad z\rightarrow\infty\,,\quad|\arg(z)|<\frac{3\pi}{4}
$$
it is possible to prove the point convergence of the series (\ref{series}) in the wedge $|\arg(s-\mu/\sigma^2)|<3\pi/4$. In the region $|\arg(s-\mu/\sigma^2)|>3\pi/4$ the asymptotic behavior is given by
$$
\mathrm{erf}(z)\sim -1-\frac{e^{-z^2}}{\sqrt{\pi}z}\,,\quad z \rightarrow \infty\,\quad |\arg(z)|>\frac{3\pi}{4}\,.
$$
In order to have point convergence in this case for (\ref{series}) we have to demand (\ref{hyperbola})
$$
\mathrm{Re}\Big(\frac{\sigma^2}{2}s^2-\mu s-\log \nu_0\Big)=0\,.
$$
The divergence of the Laplace transform on the hyperbola defined by this condition can be absorbed in the singularities of the function
\begin{eqnarray*}
F_{*\textrm{sing}}(s)=\sum_{p=0}^\infty \nu_0^{-p} \exp\left(p\Big(\frac{\sigma^2 s^2}{2}-\mu s\Big)\right)=\frac{1}{1-\exp\Big(\frac{\sigma^2 s^2}{2}-\mu s-\log\nu_0\Big)}
\end{eqnarray*}
in the region $|\arg(s-\mu/\sigma^2)|>3\pi/4$ (see Fig. \ref{Fig:sing}). These singularities control the asymptotics of the Gaussian approximation to the entropy as discussed in Section \ref{gaussian}. When the projection constraint is taken into account it is straightforward to identify the function $F_{\textrm{sing}}$ that encodes the singularities of the Laplace transform of the Gaussian approximation to the entropy
\begin{eqnarray*}
F_{\textrm{sing}}(s)=\sum_{q=1}^\infty \frac{\nu_0^{-2q}}{\sqrt{q}} \exp\left(2q\Big(\frac{\sigma^2 s^2}{2}-\mu s\Big)\right)=\mathrm{Li}_{\textrm{\textonehalf}}\left(\exp\Big(\sigma^2s^2-2\mu s-2\log\nu_0\Big)\right)\,.
\end{eqnarray*}
We end this appendix by giving an additional way to compare the exact behavior of the entropy and the Gaussian approximation. The Laplace transform inversion formula tells us that given the Laplace transform of a function $\mathcal{L}(f;s)$ it is possible to recover $f$ by using the inversion formula
$$
f(a)=\frac{1}{2\pi i}\int_{x_0-i \infty}^{x_0+i \infty} \mathcal{L}(f;s) e^{a s}\mathrm{d}s\,.
$$
The integration contour can be taken to be a line $\textrm{Re}(s)=x_0$, parallel to the imaginary axis leaving all the singularities of the integrand to its left. The Laplace transform of the exact entropy is
$$
\mathcal{L}(\exp S_*,s)=\frac{1}{1-2\sum_{k=1}^\infty\exp(-s\sqrt{k(k+2)})}\,.
$$
If we now compare the Laplace transforms $\mathcal{L}(\exp S_*,\pi\gamma+i y)$ and $\mathcal{L}(\exp S^{\textrm{gauss}}_*,\pi\gamma+i y)$ by plotting their absolute values as functions of $y$ we get the result shown in Fig. \ref{Fig:compare}.
As it can be seen there is a remarkable agreement between both plots for small values of $y$, however this agreement disappears for larger values. This is due to the fact that the real parts of the poles of the Laplace transform of the Gaussian approximation do not accumulate because they are located on hyperbolas as explained in this Appendix and in Section \ref{gaussian}.

\end{document}